\documentclass[usenatbib]{mn2e}
\usepackage{graphicx}
\usepackage{amssymb}
\usepackage{amsmath}
\usepackage{times}
\usepackage{aas_macros}

\bibpunct[,]{(}{)}{;}{a}{}{,}

\begin{document}

\title[Multifractal and lacunarity analysis  of the galaxies of  SDSS Data Release 9]{Multifractal analysis and lacunarity spectrum of the galaxies of the ninth Sloan Digital Sky Survey (SDSS) data release}
\author[C. A. Chac\'on-Cardona, R. A.  Casas-Miranda]{C. A. Chac\'on-Cardona$^{1,2}$, R. A.  Casas-Miranda$^1$\\
		$^1$Departamento de F\'{\i}sica, Universidad Nacional de Colombia\\
		$^2$Facultad Tecnol\'ogica, Universidad Distrital Francisco Jos\'e de Caldas\\}
		
\date{Accepted ---. Received ---; in original form ---}

\pagerange{\pageref{firstpage}--\pageref{lastpage}} \pubyear{2012}

\maketitle
\label{firstpage}

\begin{abstract}

In this work, we develop a statistical analysis of the large-scale clustering of matter in the Universe from the fractal point of view using galaxies from the Ninth Sloan Digital Sky Survey (SDSS) Data Release (DR9). From the total set of galaxies, a magnitude-limited sample of galaxies with redshifts in the range $0 < z < 0.15$ was created. The sample covers the largest completely connected area of the celestial sphere within the catalogue, with limits in right ascension of  $120 ^\circ < \alpha < 240 ^\circ $   and declination $0 ^\circ < \delta < 60 ^\circ$, which is a region that includes the largest galactic samples that have been studied from the fractal viewpoint to date. The sample contains 164,168 galaxies.

Using the sliding-window technique, the multifractal dimension spectrum and its dependence on radial distance are determined. This generalisation of the concept of fractal dimension is used to analyse large-scale clustering of matter in complex systems. Likewise, the lacunarity spectrum, which is a quantity that complements the characterisation of a fractal set by quantifying how the set fills the space in which it is embedded, is determined.

Using these statistical tools, we find that the clustering of galaxies exhibits fractal behaviour that depends on the radial distance for all calculated quantities. A transition to homogeneity is not observed in the calculation of the fractal dimension of galaxies; instead, the galaxies exhibit a multifractal behaviour whose dimensional spectrum does not exceed the physical spatial dimension for radial distances up to 180 Mpc/h from each centre within the sample. Our results and their implications are discussed in the context of the formation of large-scale structures in the Universe.

\end{abstract}


\section{Introducci\'on}

The standard cosmological principle states that the cosmos is homogeneous and isotropic on large scales \citep{Longair:2008}, such that the observed inhomogeneities are only perceived locally and should vanish on sufficiently large scales, at distances on the order of hundreds of Mpc/h \citep{2009arXiv0907.2127G, citeulike:2502783}. The majority of the developments in modern cosmology use the cosmological principle as the fundamental hypothesis from which the observations and theoretical developments are contrasted and interpreted. The success of the solution to the Einstein field equations for a universe consisting of an ideal gas of galaxies, found by Friedmann, Robertson and Walker, makes the corresponding metric the representation of the space-time structure of the Universe that is largely accepted in cosmology for the study of large-scale matter clustering.
The  $\Lambda$CDM model strongly relies on the standard cosmological principle because it introduces the cosmological constant as a remedy for the problem of the acceleration of the Universe, which is confirmed by observations of Type Ia supernovae. However, the model excludes possible solutions that do not assume homogeneity, such as models based on the metric of Tolman and Bondi \citep{1992ApJ...388....1R, 2008GReGr..40..451E}. 

Despite the success of physical models based on the standard cosmological principle, there are still many unresolved fundamental questions regarding the formation of structure in the Universe \citep{peacock1999cosmological}. For example, is the cosmological principle supported by the observations of galaxies in the most recent catalogues? If so, then what is the distance scale at which the transition to homogeneity is observed? Some research groups report that galaxies and other astrophysical objects, such as quasars, are grouped in highly structured hierarchical patterns that exhibit properties of self-similarity (scale invariance) and a fractal dimension smaller than the physical space dimension \citep{2012arXiv1201.5554R}. Similar results are found for the anisotropies of the cosmic background radiation \citep{2011JPSJ...80g4003K}. In such works, a revision of the cosmological principle is proposed because, in a universe with fractal clustering of galaxies, models based on homogeneity and isotropy would not be consistent with observations. Therefore, new methods of interpreting galaxy clustering information from the perspective of statistical mechanics should be developed \citep{citeulike:2502783}. Furthermore, other research groups (e.g., \citep{2009MNRAS.399L.128S}) claim to have found the scale at which the cosmological principle is valid. In their multifractal analysis of the Sixth SDSS Data Release (DR6), they find a transition to homogeneity at approximately 70 Mpc/h. Similarly, analyses of galactic catalogues with information from the southern hemisphere conclude Ð using a method based on the correlation dimension - that the transition to homogeneity occurs on the same spatial scale \citep{Scrimgeour:2012wt}. 

In this article, we present calculations of the multifractal dimension and the corresponding lacunarity spectrum based on the generalised correlation integral for all galaxies in the Ninth SDSS Data Release (DR9). We use the same methodology that was applied to dark matter halos from the Millennium Simulation, an N-body simulation with gravitational interaction. For the simulated halos, we found a transition to homogeneity at radial distances between 100 Mpc/h and 120 Mpc/h, and the lacunarity spectrum confirmed such transition \citep{2012arXiv1209.2637C}.

This paper is organised as follows. In section 2, the mathematical foundations involved in the determination of multifractality and the lacunarity spectrum of galaxy clustering is reviewed. Section 3 contains a summary of the characteristics of DR9 and of the primary magnitude-limited sample of galaxies that we use. In section 4, the calculations of multifractality and lacunarity are discussed for the case of galaxy clustering up to a redshift of 0.15, which corresponds to a maximum radial distance of approximately 600 Mpc/h, where the effect of space-time curvature is negligible. In section 5, the results of multifractality and lacunarity are discussed, and the conclusions are presented in section 6.

\section{Theoretical Formalism}

\subsection{Correlation Dimension and Lacunarity}

Fractal analysis of the clustering of galaxies is based on the idea of generalising the dimension of a metric space. This concept, developed by Felix Hausdorff in \citep{springerlink:10.1007/BF01457179}, allowed Benoit Mandelbrot to define a fractal as a set with Hausdorff dimension strictly exceeding its topological dimension (\citep{mandelbrot1983fractal}, where the topological dimension is the number of independent directions that can be taken around an element belonging to a given set. It is possible to determine the Hausdorff dimension for finite sets using computational algorithms, such as the minimum spanning tree (MST). However, the correlation dimension $D_2$, is a very useful tool for measuring fractality on large scales  \citep{martinez2002statistics}.

The definition of the correlation dimension is based on the integral of the correlation function $C_{2}$, which is a function that measures the number of neighbours found on average within a certain radius $r$ around a given centre. In agreement with the notation in the literature used by \citet{2008MNRAS.390..829B}, it is defined as:

\begin{equation}\label{C2}
 C_{2}(r)=\frac{1}{\mathit{NM}}\sum_{i=1}^M n_{i}(r),
\end{equation} where $N$ is the total number of particles within the distribution, $M$ is the number of particles used as centres, and the sum is taken over the subset of chosen centres. Here $n_{i}(r)$,  is the number of particles within a radial distance $r$  with respect to a particle $i$ chosen as the centre. This expression is defined as:

\begin{equation}\label{ni}
 n_{i}(r)=\sum_{j=1}^N \Theta(r-\mid \boldsymbol{x_i-x_j} \mid),
\end{equation} where the sum is taken over all the particles in the sample. The coordinates of each particle in our three-dimensional space are denoted by $\boldsymbol{x_j}$, and $\Theta$ is the Heaviside function, which is defined as $\Theta(x) = 0$ for $x < 0$ and $\Theta(x)=1$ for $x \geqslant 0$. The number of particles around each centre  is determined by counting the number of galaxies around the centre, i.e., the galaxies found within a sphere of radius $r$ around the centre.
 
From Equation (\ref {C2}), the correlation dimension is expressed in the following form:

\begin{equation}\label{D2}
 C_{2}(r) = F r^{D_{2}}
\end{equation} where  $F$ is called the pre-factor of the power law and  is the correlation dimension of the set under study. Thus, the correlation dimension is calculated as the derivative:
 
\begin{equation}
  D_{2}=\frac{d\log C_{2}(r)}{d\log r}
\end{equation}

By expressing the correlation integral as a power law, in addition to the multifractal dimension, the set is fully characterised by the proportionality constant $F$ that accompanies the function. The pre-factor $F$ which is related to the average distance between the nearest neighbours, permits defining the lacunarity, which is a property that determines how the fractal set fills the space. Defined by  \citet{PhysRevE.56.112} via the second-order variability factor for the pre-factor $F$, the lacunarity is expressed as:

\begin{equation}
\Phi =\frac{\bigl\langle(F - \bigl\langle F \bigr\rangle)^{2}\bigr\rangle}{\ \bigl\langle F \bigr\rangle^{2}}=\frac{\bigl\langle F^{2}\bigr\rangle}{\bigl\langle F\bigr\rangle^{2}}-1.
\end{equation}

Whereas lacunarity was defined as a complementary concept to the fractal mass-radius dimension, this concept can be applied to other definitions of fractal dimension, as shown in \citep{martinez2002statistics}; The correlation dimension and its corresponding lacunarity are sufficient to characterise sets for which these quantities do not depend on the scale, as discussed in the majority of studies that demonstrate fractal behaviour, at least on small scales. Therefore, in the case of scale dependence of fractal quantities we must extend the statistical analysis by using the formalism of the multifractal dimension \citep{2004PhyA..341..215N, citeulike:2502783}.

\subsection{Generalised Fractal Dimension}

Based on the correlation integral  $C_{2}(r)$, given Equation (\ref{C2}), the generalised correlation integral can be defined via the relation:

 \begin{equation}
 C_{q}(r)= \frac{1}{\mathit{NM}}\sum_{i=1}^M [n_{i}(r)]^{q-1},
\end{equation} where $M$ is the number of centres, $N$ is the number of particles included in the sample,  $n_{i}(r)$  is the same expression defined in Equation (\ref {ni}), and $q$ is the structure parameter, which corresponds to an arbitrary real number. From the generalised correlation integral, it is possible to make a power series expansion of  $log(r)$ as described in \citet{1997Chaos...7...82P}, and thus be able to calculate the multifractal dimension and the lacunarity spectrum directly:

 \begin{equation}
log \left[ C_{q}(r)^{1/(q-1)}\right] =D_{q}log(r)+log(F_{q})+O \left( \frac{1}{log¨}\right).
\end{equation} By retaining only the first two terms on the right side of the equation, we obtain the relationship between the generalised correlation integral and the multifractal dimension:

\begin{equation}
C_{q}(r)^{1/(q-1)}=F_{q} r^ {D_{q}}.
\end{equation}

The generalised fractal dimension and the generalised lacunarity can be defined in the same manner by which the authors of  \citet{PhysRevE.56.112}, defined these concepts for the mass-radius fractal relation:

\begin{equation}
  D_{q}=\frac {1}{(q-1)} \frac{d\log C_{q}(r)}{d\log r},
\end{equation} 

Likewise, the corresponding generalised lacunarity based on the pre-factor $F_q$  for each structure parameter for $q$ is defined by:

\begin{equation}
\Phi _{q}=\frac{\bigl\langle(F_{q} - \bigl\langle F_{q} \bigr\rangle)^{2}\bigr\rangle}{\ \bigl\langle F_{q} \bigr\rangle^{2}}=\frac{\bigl\langle F_{q}^{2}\bigr\rangle}{\bigl\langle F_{q} \bigr\rangle^{2}}-1.
\end{equation}

In the fractal dimension spectrum, if  holds for any  $q\neq q'$, $D_{q} = D_{q'}$  then the distribution is said to be a homogeneous fractal (monofractal). $q \geqslant 1$, $D_q$ characterises the scaling behaviour of the system in high-density environments, i.e., clusters or super-clusters of galaxies. Conversely, for values $q < 1$,  characterises the scaling behaviour in low-density environments, i.e., vacuum regions \citep{2007ApJ...658...11G, 2009MNRAS.399L.128S}.  When the distribution of galaxies exceeds the transition to homogeneity, all the values of the multifractal dimension should approach the physical dimension of the space  and the lacunarity spectrum must approach zero  on the same radial distance scale $r$.

Next, we apply the concepts of generalised fractal dimension and lacunarity to the distribution of galaxies from the latest update of the SDSS catalogue. 

\section{Galaxies from the ninth data release of the SDSS catalogue}

The SDSS galactic catalogue is the largest astronomical database accumulated by humankind to date. Its ninth edition contains information regarding at least 1231,051,050 astrophysical objects, including spectral information for 668,054 stars, 1,457,002 galaxies and 228,468 quasars and their respective locations in the celestial sphere over an area of 14,555 square degrees \citep{2012arXiv1207.7137S}. In its ninth data release, the catalogue has two clearly defined regions without any information gaps in their interiors,  therefore in these regions it is possible to calculate the fractal dimension without the risk of losing accuracy. From these two regions, we chose the region with the largest connected area to have a more representative sample of the galaxy catalogue; the region covers a total area of 7,200 square degrees.

\begin{figure}
\includegraphics[width=1 \linewidth,clip]{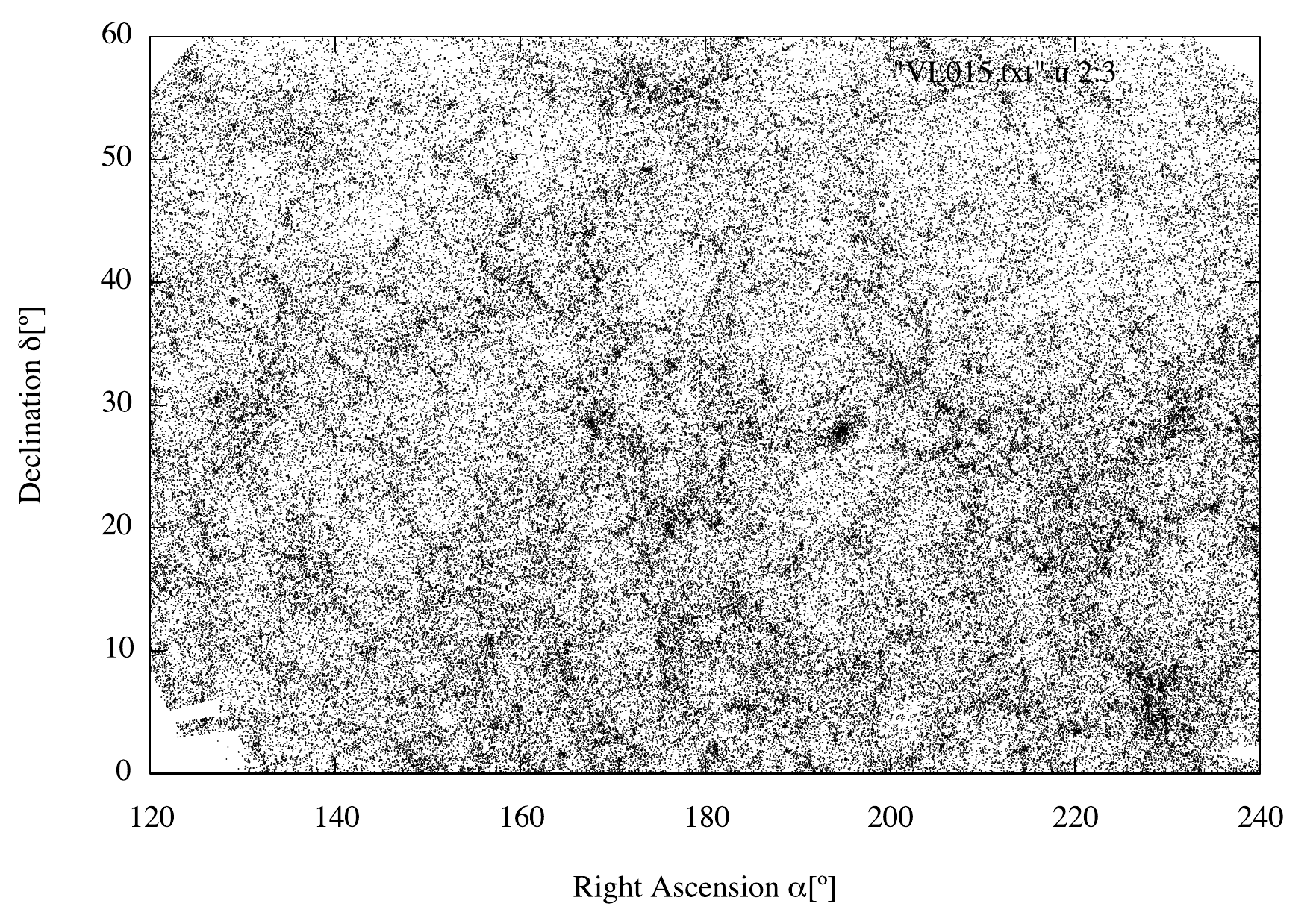}
 \caption{Distribution on the celestial sphere of galaxies located in the largest area completely covered by SDSS. The region is bounded by equatorial coordinates of right ascension $120 ^\circ < \alpha < 240 ^\circ $ and declination $0 ^\circ < \delta < 60 ^\circ$ }
	\label{fig: SDSS_eq}
\end{figure}

The information contained in this data release has been the subject of two different targeting algorithms, one for galaxies with redshifts in the range $ 0.15 < z < 0.4 $, called LOWZ, and a second sample of galaxies at higher redshifts, with$ 0.4 <  z  <  0.8 $ , called CMASS. While we could use either of these two samples for the fractal characterisation of galaxy clustering, we wanted to avoid erroneously introducing homogenising effects arising from the determination of distances according to the standard cosmological model, as pointed by Ribeiro \citep{2005A&A...429...65R} and the group \citep{citeulike:2502783}. Using Hubble's law, which is an experimental fact valid for astrophysical object localisation at low redshift ($z \ll 1$), we selected from DR9 galaxies with redshifts in the range $2.3\times10^{-3} < z < 0.15$, corresponding to distances greater than 10 Mpc/h, where the effect of peculiar velocities is small, and less than 600 Mpc/h, where the difference between the distance determined using the Hubble law and the co-moving distance determined according to the standard cosmological model does not exceed $4\%$.. In this region, the targeting algorithm most suitable for reliable detection of galaxies is the Main Sample of galaxies, a Petrosian r-band magnitude-limited galactic sample, whose selection process is described in detail by \citet{2002AJ....124.1810S}

\subsection{Main sample of galaxies}

The SDSS redshift catalogue includes galaxy photometric data corrected for galactic extinction, where reddening calculations are performed according to the maps of infrared emission from galactic dust described in \citet{1998ApJÉ500..525S}. All information necessary for fractal analysis is provided by the catalogue in CASJOBS, which is accessed using SQL (Syntax Query Language). First, galaxy candidates are separated from stars. Then, two cuts in apparent magnitude are applied for the Petrosian r spectral band, which is located at approximately 6165 . The apparent magnitude for acceptable targets is in the interval $ 15 < m_r < 17.77 $, where the first limit indicates that those targets with apparent magnitude less than 15 are discarded, i.e., very bright objects that cause contamination in the sample are discarded. We also discard objects with Petrosian radii, which contain half the total flux of the galaxy, greater than two arcseconds $(r_p > 2" )$. The second limit indicates that it is not possible to detect objects that are fainter in apparent magnitude than the limit of 17.77. In addition, it is necessary to apply a condition on the average surface brightness determined according to the Petrosian radius; the most significant cut is the cut that includes targets with values $\mu_{50} \leqslant 23 mag/arcsec^2$.

\begin{figure}
\includegraphics[width=1 \linewidth,clip]{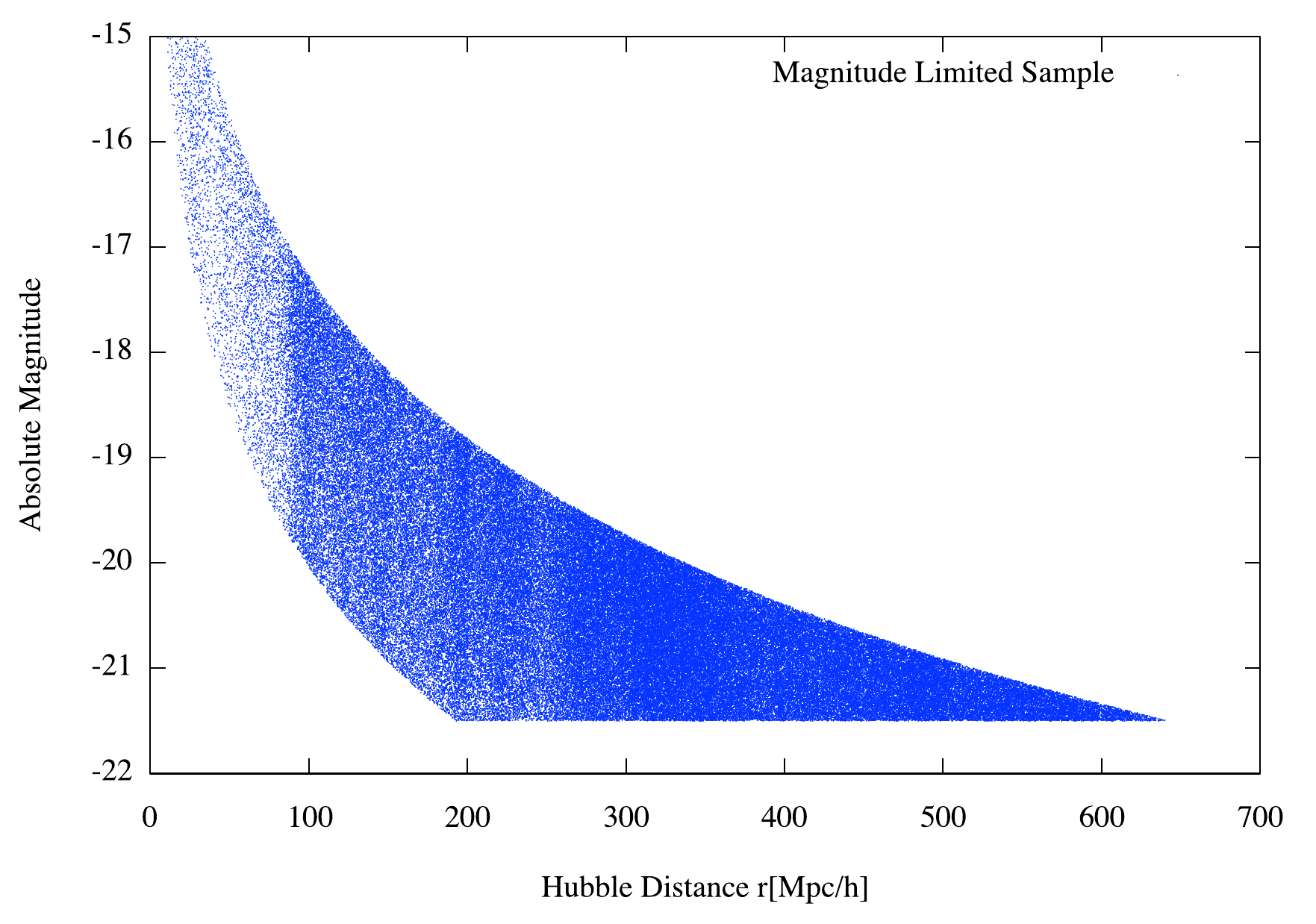}
 \caption{Magnitude-radial distance diagram for the primary sample of galaxies from DR9 used for the multifractal calculations. The sample is composed of 164,168 galaxies with redshifts in the range $2.3\times10^{-3} < z < 0.15$.}
	\label{SDSS}
\end{figure}

Given the apparent magnitude limits, to obtain a sample that is not incomplete because of the effects of selection by brightness, the procedure includes the determination of the volume that limits the sample according to the range of distances over which one wishes to work. Because our interest is measuring fractality without introducing any a priori cosmological model, with distances in the range $10 Mpc/h \leqslant r \leqslant 600 Mpc/h$ the absolute magnitude of the galaxies in the sample are in the range   $ -15 \geqslant M \geqslant -21.5$,  as shown in Figure  \ref{SDSS}.

Finally, for the volume-limited sample, we must determine the radius of the largest sphere in which fractality calculations are performed without introducing assumptions about the shape of the clustering to be analysed. The effective depth scale for our sample according to the limits in right ascension $120 ^\circ < \alpha < 240 ^\circ $, and declination $0 ^\circ < \delta < 60 ^\circ$, is determined in accord with \citep{citeulike:2502783} by the relationship:

\begin{equation}
  R_{s}= \frac{R_d sin(\delta \theta/2) }{1+ sin(\delta \theta/2) },
\end{equation}where $R_d$  corresponds to the maximum radial distance of the sample and $\delta \theta$ is

\begin{equation}
 \delta \theta = min(\alpha_2-\alpha_1,\delta_2-\delta_1)
 \end{equation}In our case the maximum radius is $R_{s}=$ 202 Mpc/h.
 
Starting with this sample, the next step involves the determination of the multifractal dimension and lacunarity spectrum up to depths of 202 Mpc/h, which corresponds to the effective depth scale of our volume-limited sample that relies on a number of available centres for the analysis of 81,435 galaxies

\section{Multifractality and lacunarity spectrum for galaxy clustering in SDSS DR9}

\begin{figure*}
	\includegraphics[width=0.45\linewidth,clip]{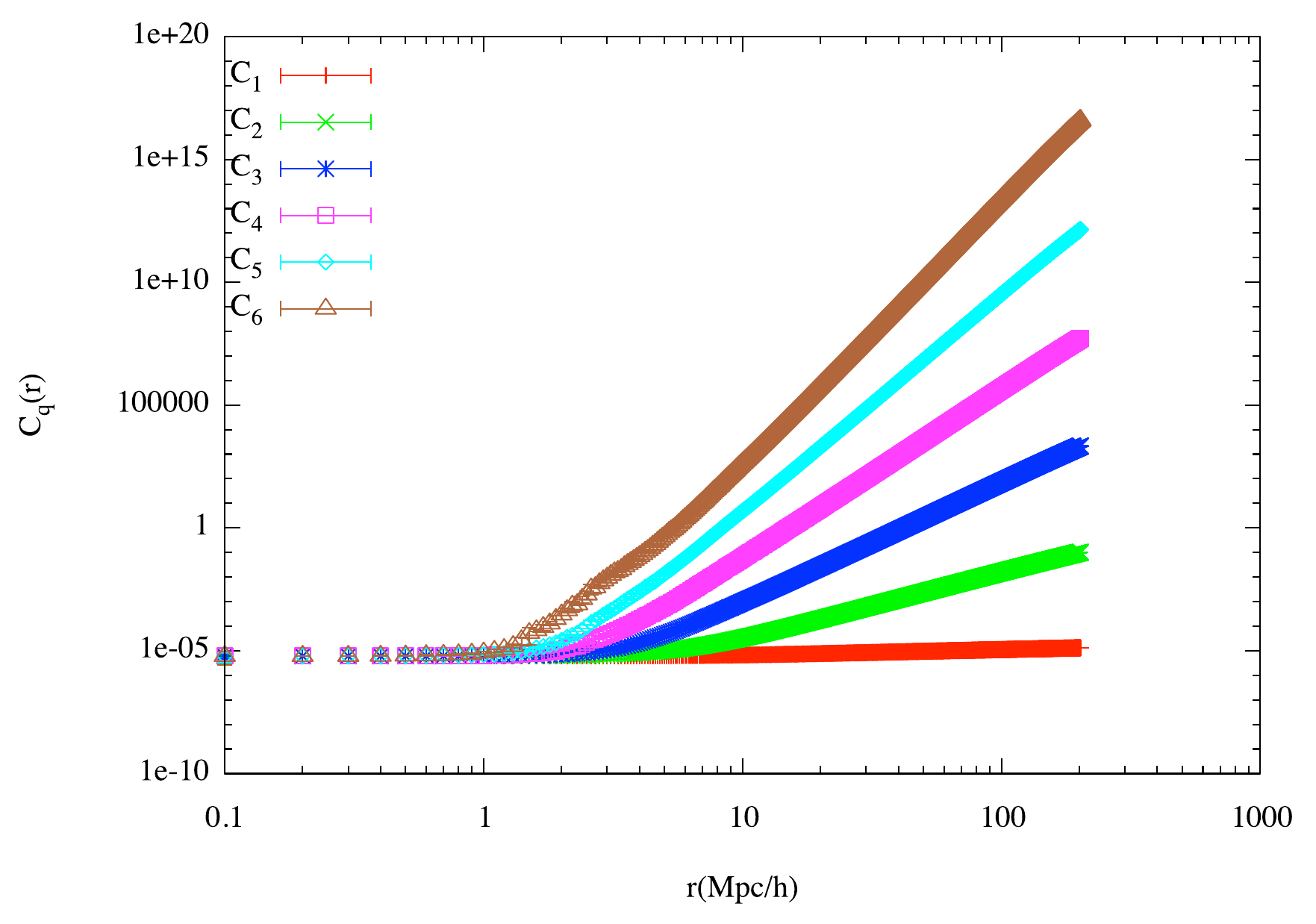}
	\includegraphics[width=0.45\linewidth,clip]{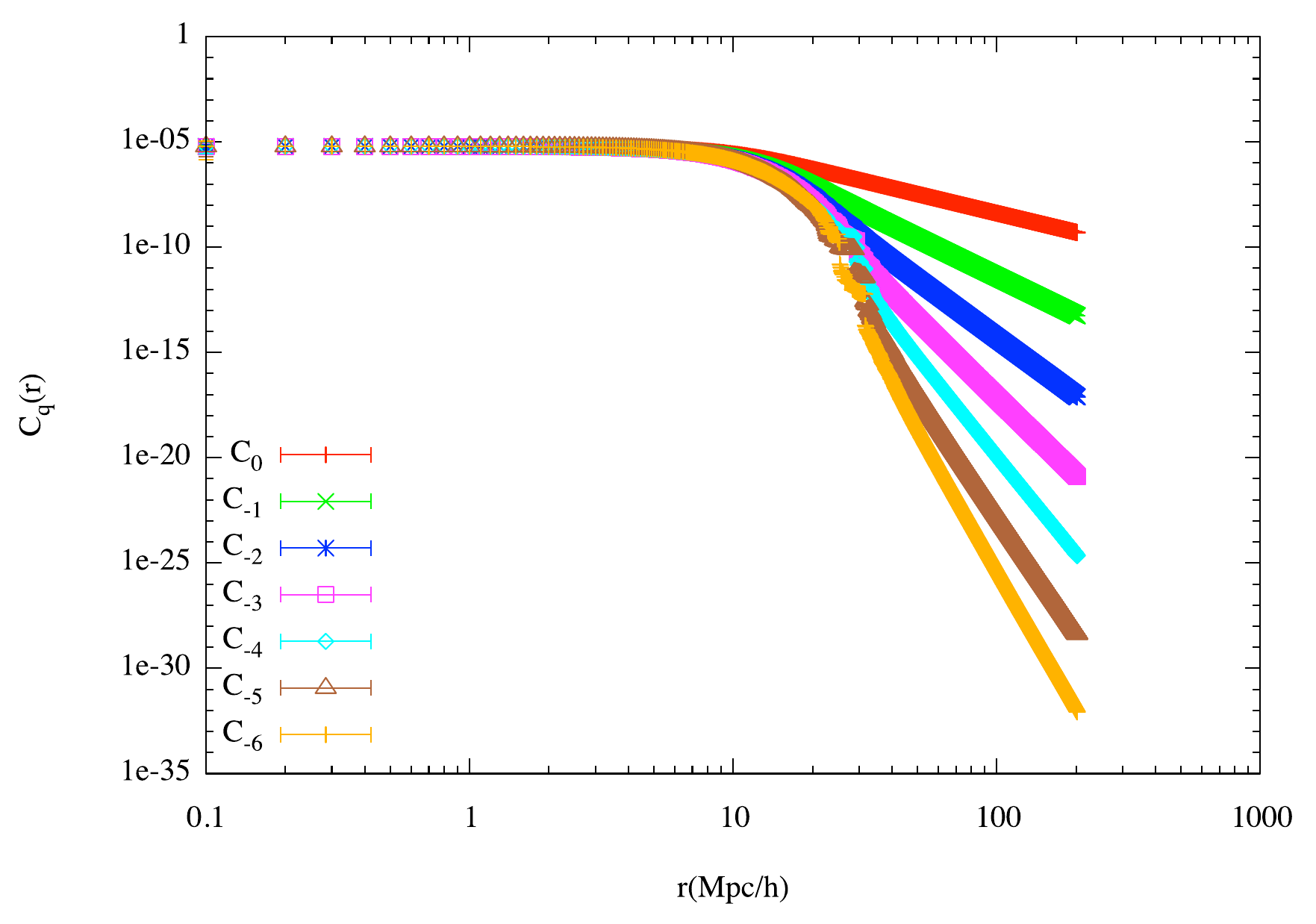}

 \caption{Generalised correlation integral  for all values of $q$ studied in this article for our volume-limited sample from DR9. Densities with $q \geqslant 1 $ appear to the left, whereas low densities,  with $q < 1$, appear on the right. All the graphs are shown in log scale}
	\label{Cqr}
\end{figure*}

In the previous section, we analysed the volume-limited sample of galaxies with redshifts in the range $2.3\times10^{-3} < z < 0.15$ . This set is composed of 164,168 galaxies for which we calculated the generalised correlation integral. We used 13 values of the structure parameter in the range $-6 \leqslant q  \leqslant 6  $ in steps of 1, which are presented in Figure \ref{Cqr}. These calculations include high-density regions, i.e., $(q \geqslant1)$, and low-density regions $(q<1)$. For $q=1$, the numerical limit is shown in the figure.

\begin{figure*}
	\includegraphics[width=0.45\linewidth,clip]{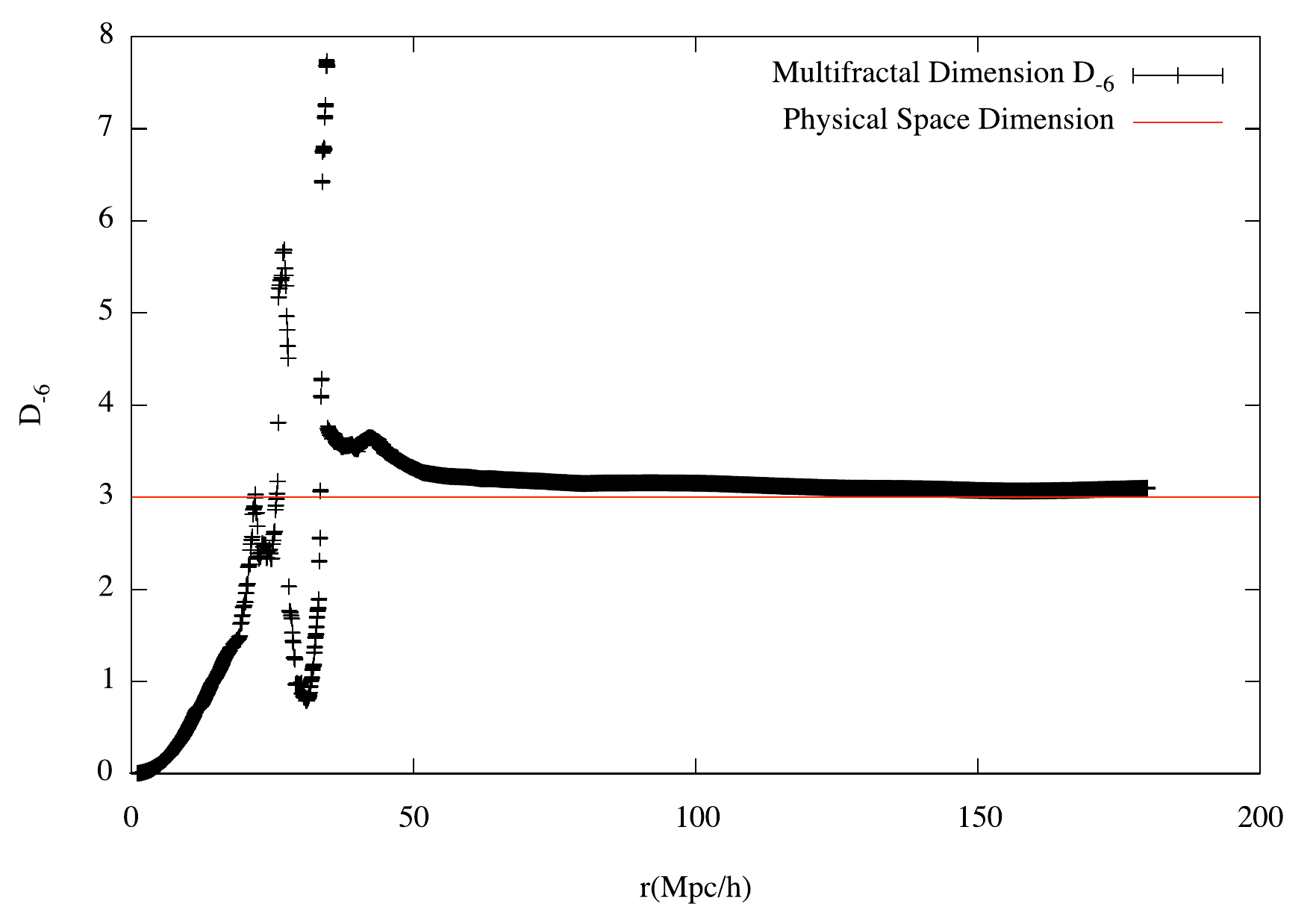}
	\includegraphics[width=0.45\linewidth,clip]{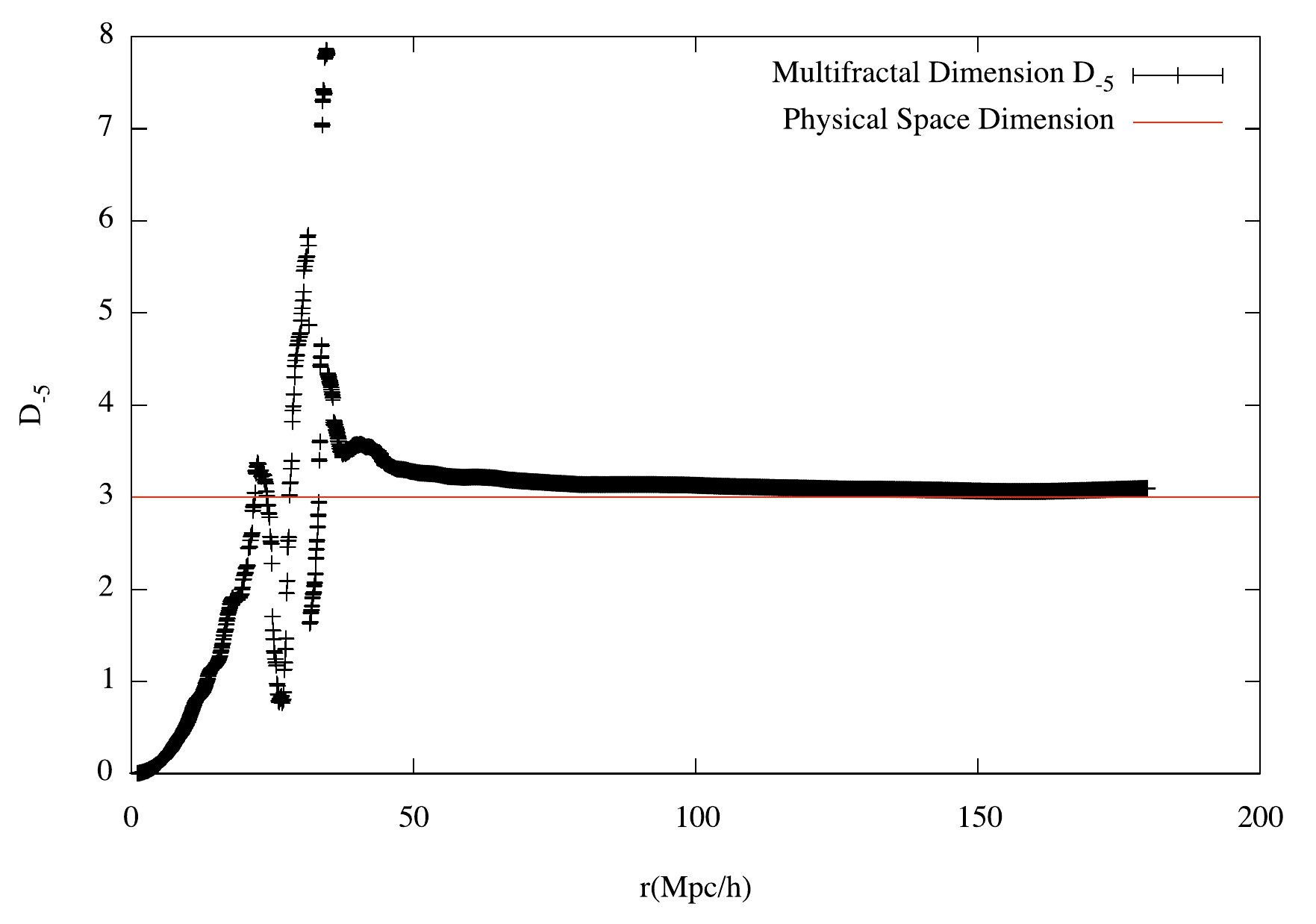}
	\includegraphics[width=0.45\linewidth,clip]{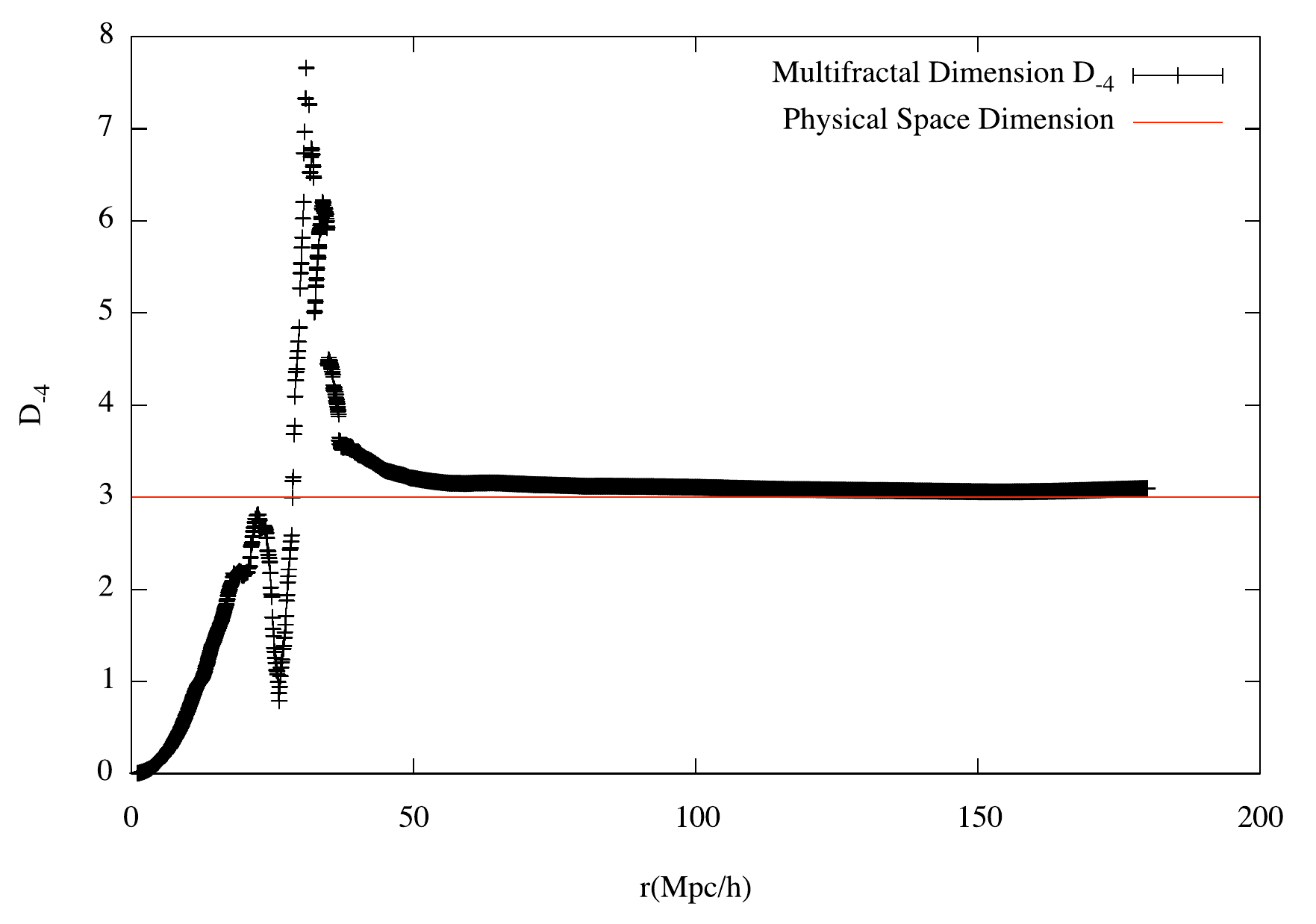}
         	\includegraphics[width=0.45\linewidth,clip]{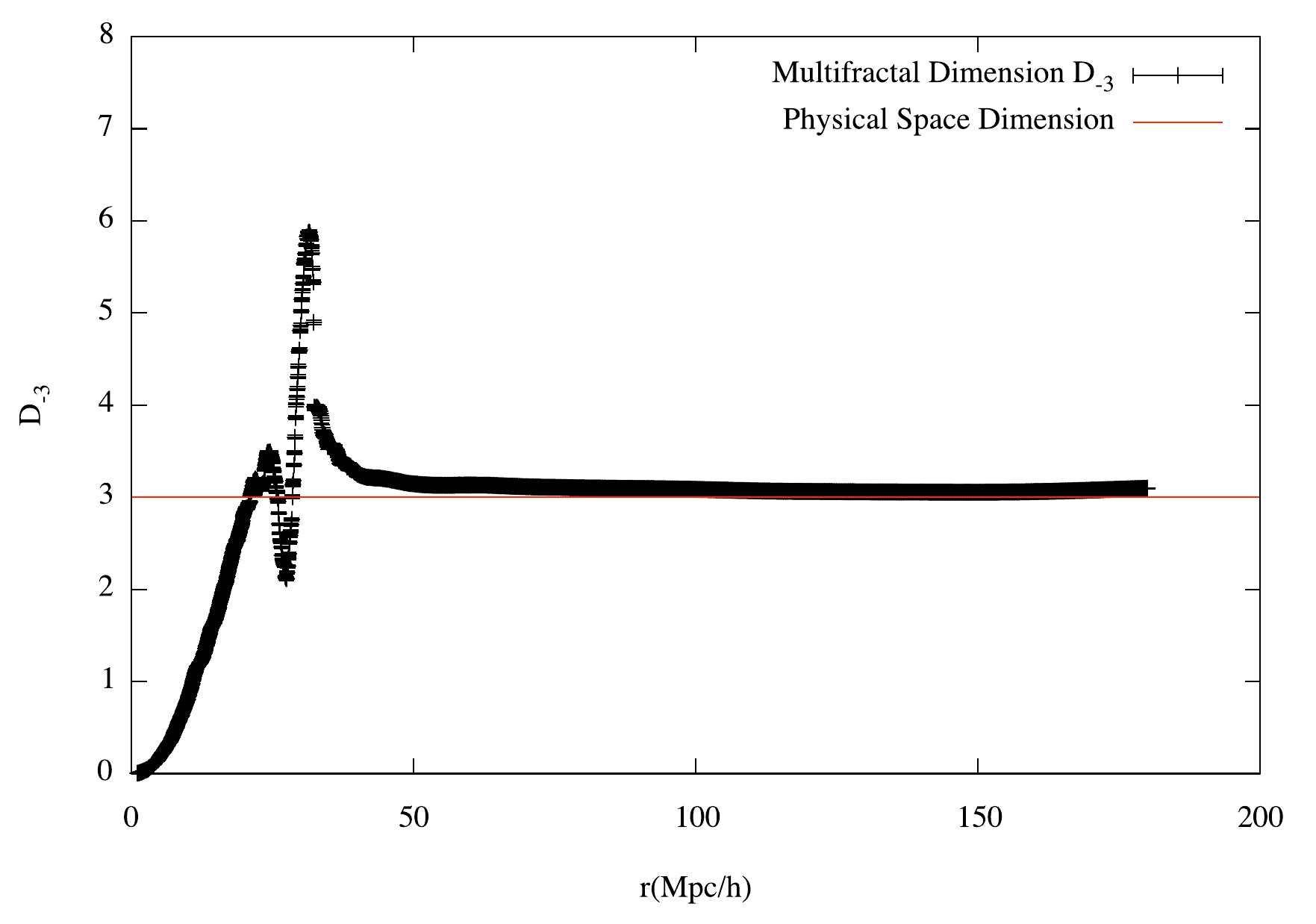}
	\includegraphics[width=0.45\linewidth,clip]{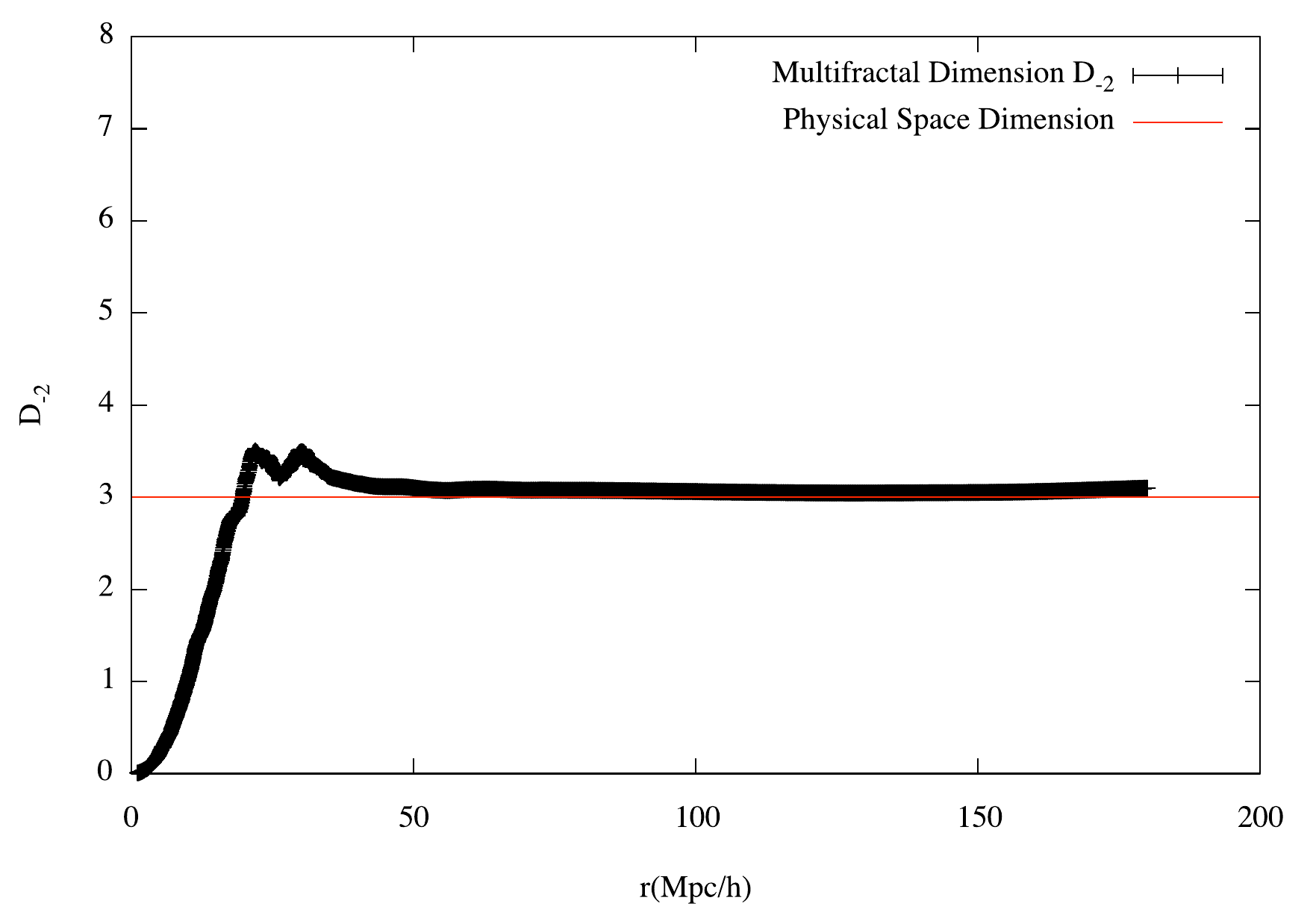}
 	\includegraphics[width=0.45\linewidth,clip]{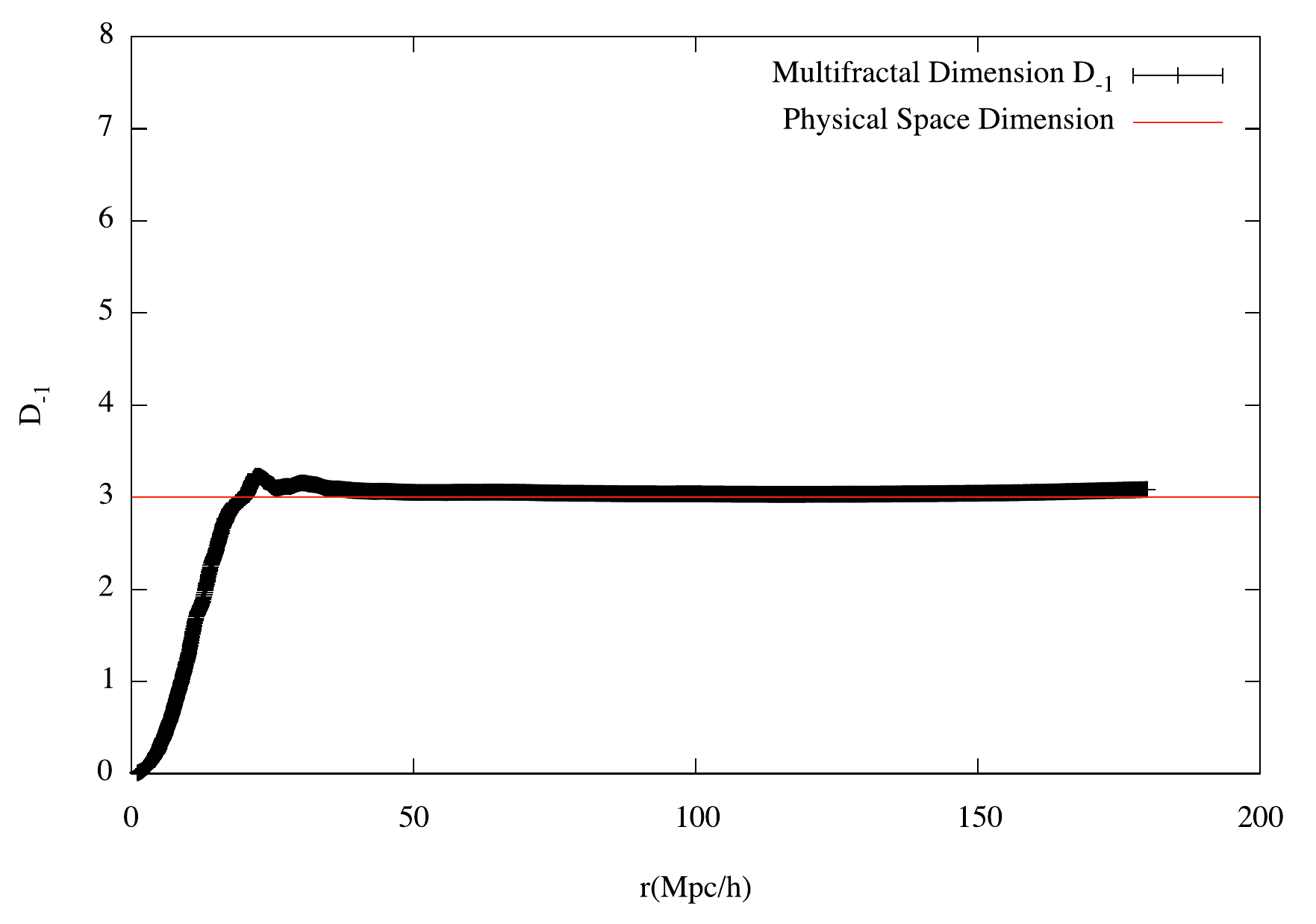}
         \includegraphics[width=0.45\linewidth,clip]{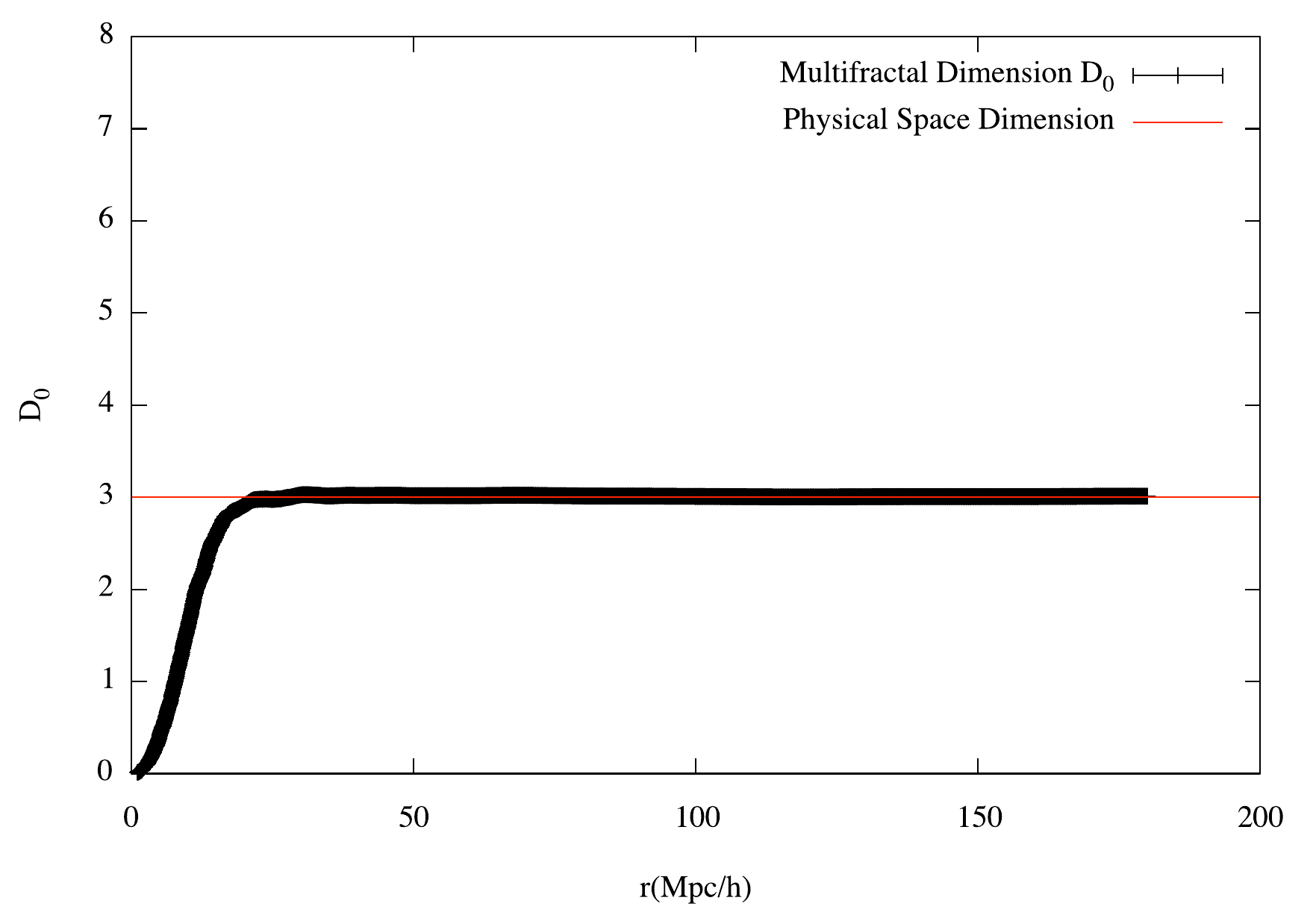}
	
	\caption{Multi-fractal spectrum  $D_q(r)$  as a function of the radial distance $r$ for low-density environments, with $?6\leqslant q \leqslant 0$, corresponding to the galaxy clustering in SDSS DR9. Error bars  $\pm1\sigma$  are indicated in the graphs. } 
	\label{Dqr1a}
\end{figure*}

The fractal dimension spectrum is determined from the generalised correlation integral using the sliding-window technique. This technique is applied to the logarithm of the correlation integral as a function of the logarithm of the radial distance. This function is approximated by straight lines using the least squares method. Below, Figure  \ref{Dqr1a} shows the results of the calculations of the fractal dimension spectrum for low-density regions with structure parameter values $q < 1$. 

\begin{figure*}
	\includegraphics[width=0.45\linewidth,clip]{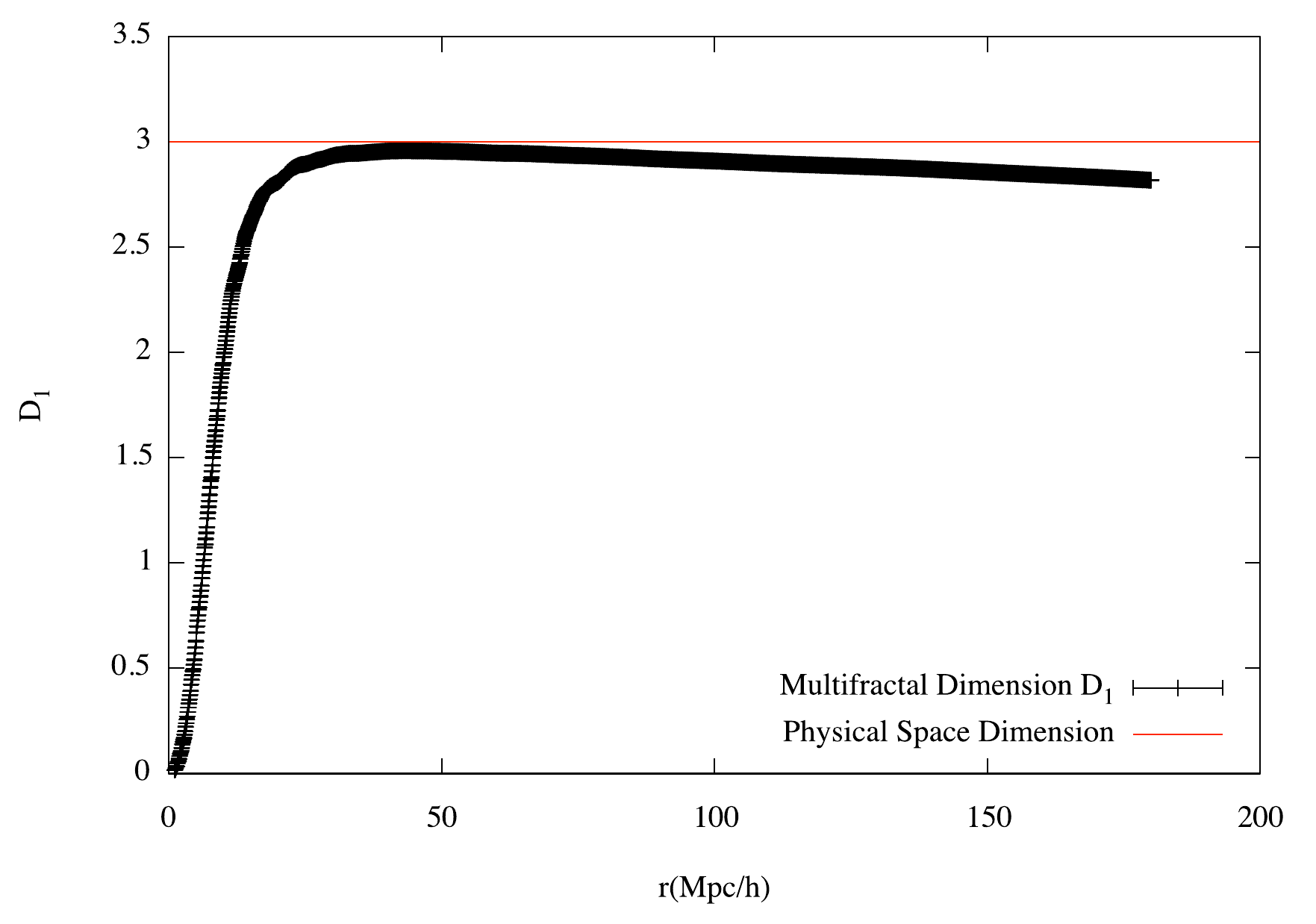}
	\includegraphics[width=0.45\linewidth,clip]{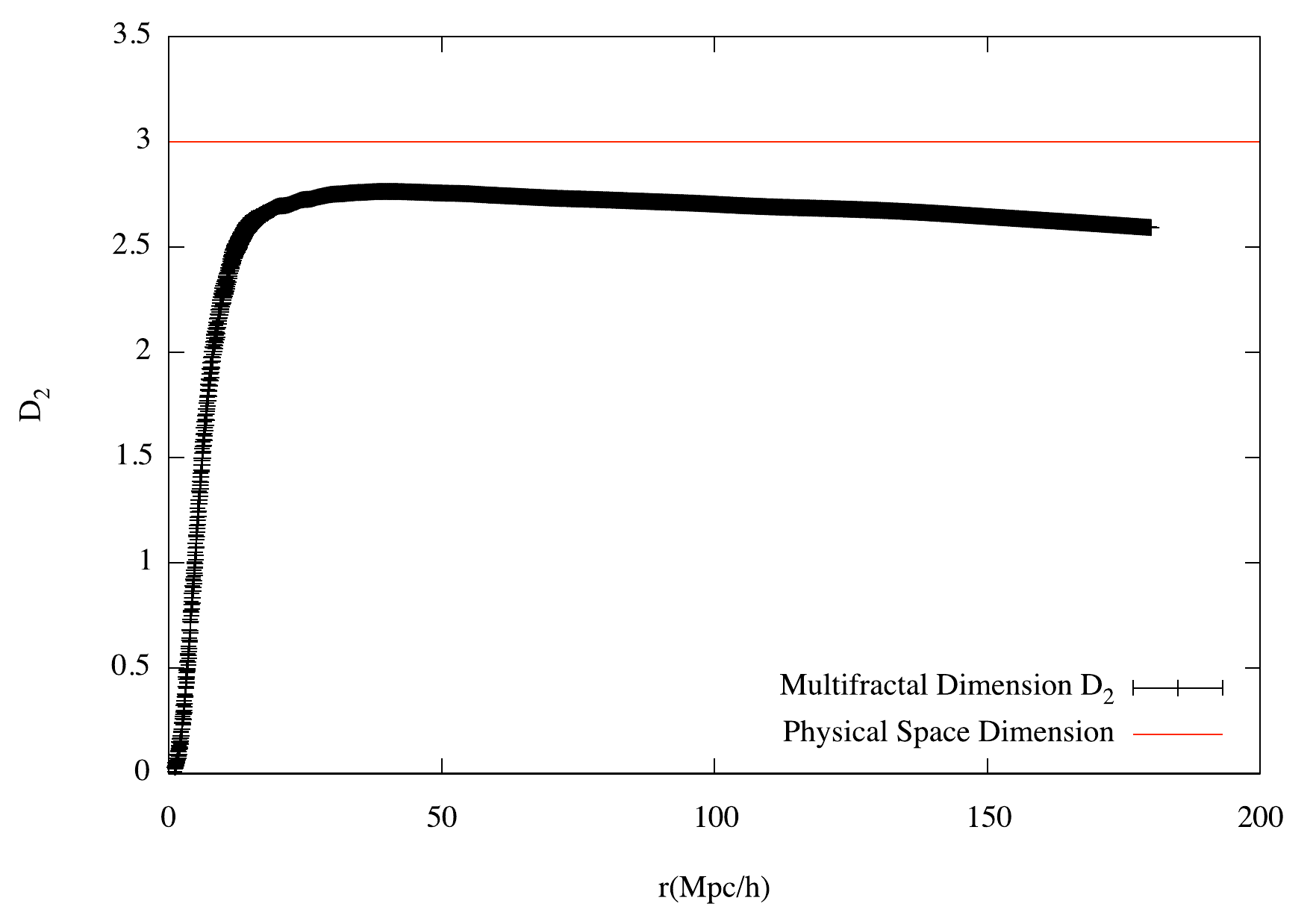}
	\includegraphics[width=0.45\linewidth,clip]{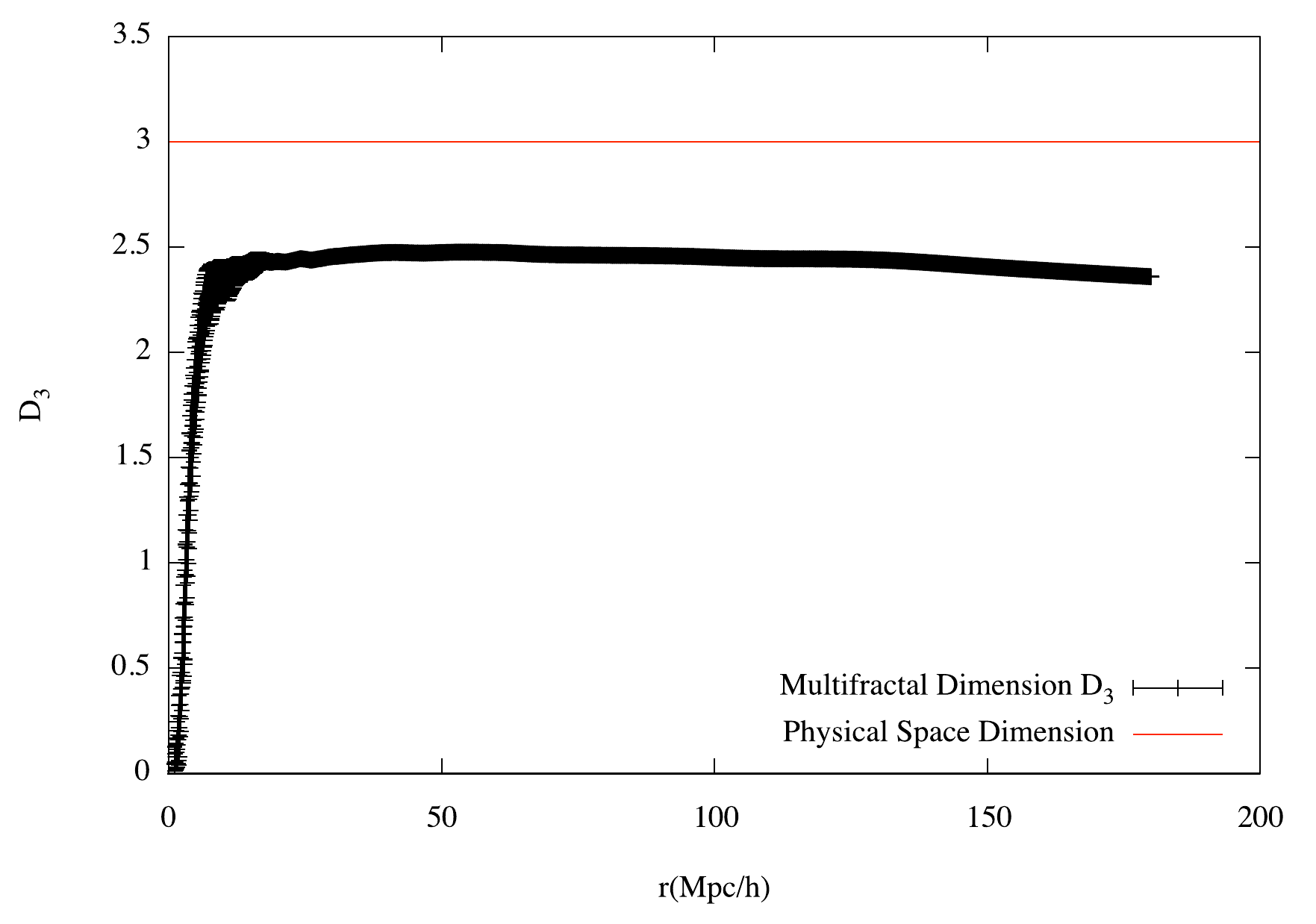}
	\includegraphics[width=0.45\linewidth,clip]{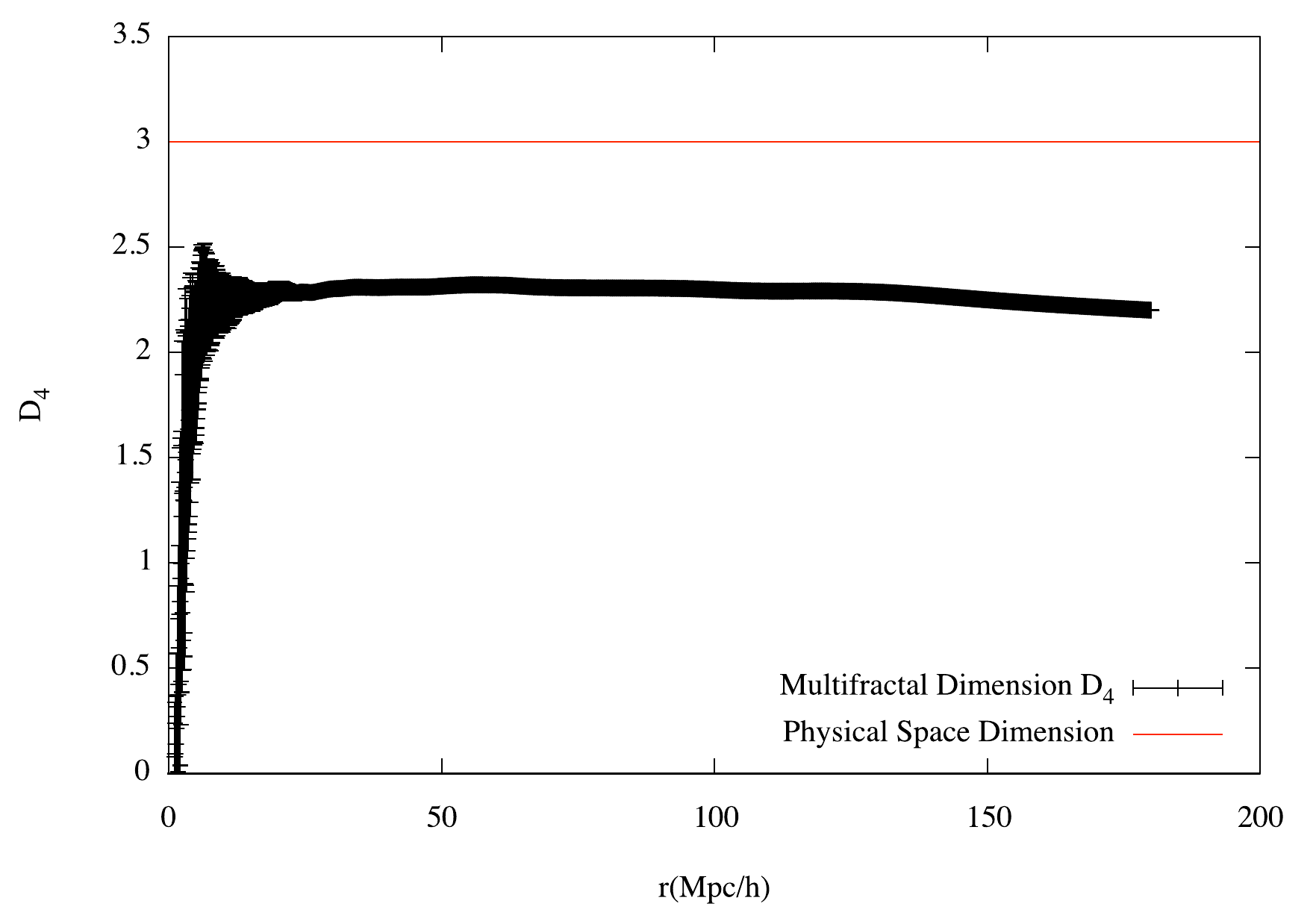}
	\includegraphics[width=0.45\linewidth,clip]{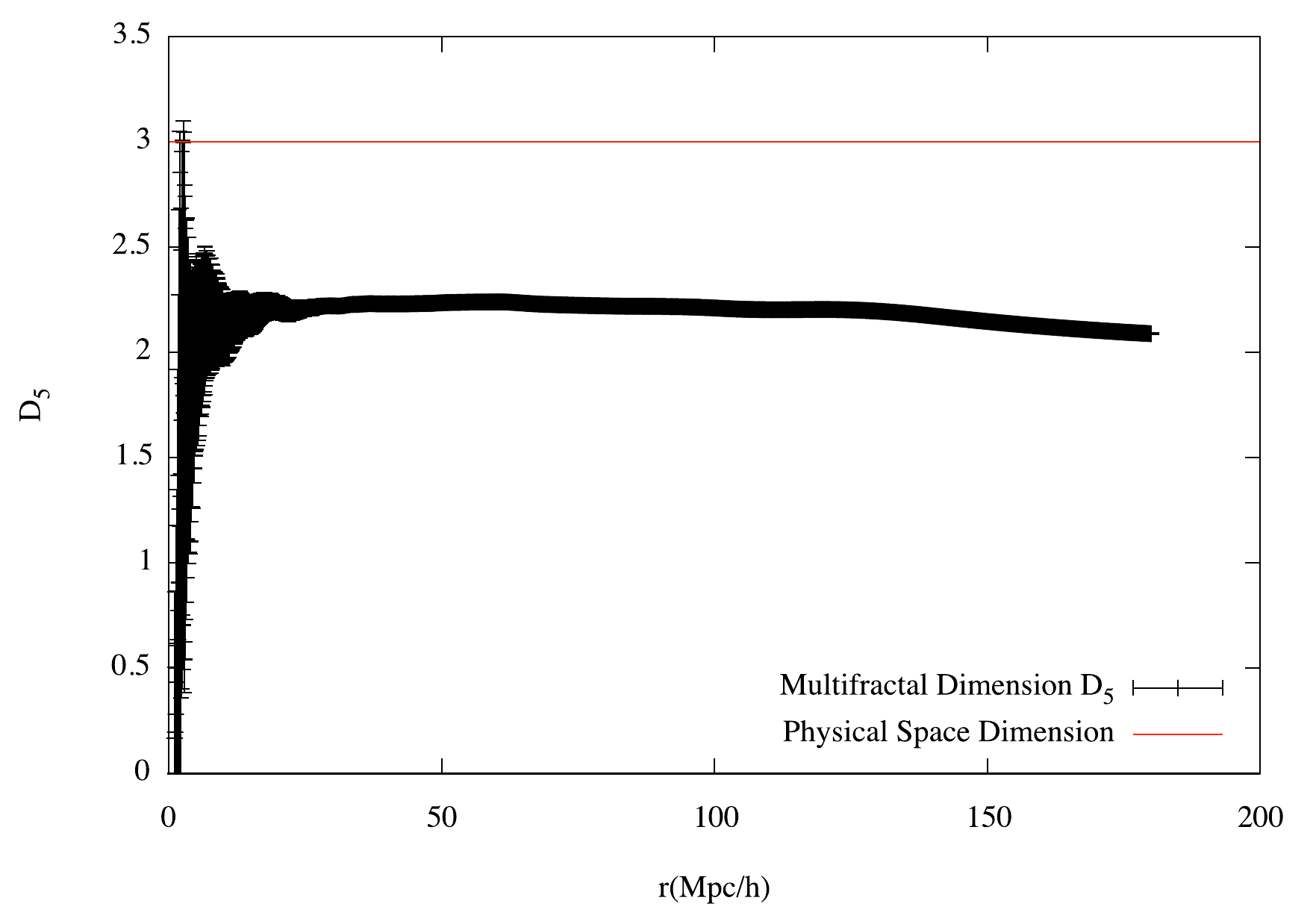}
	\includegraphics[width=0.45\linewidth,clip]{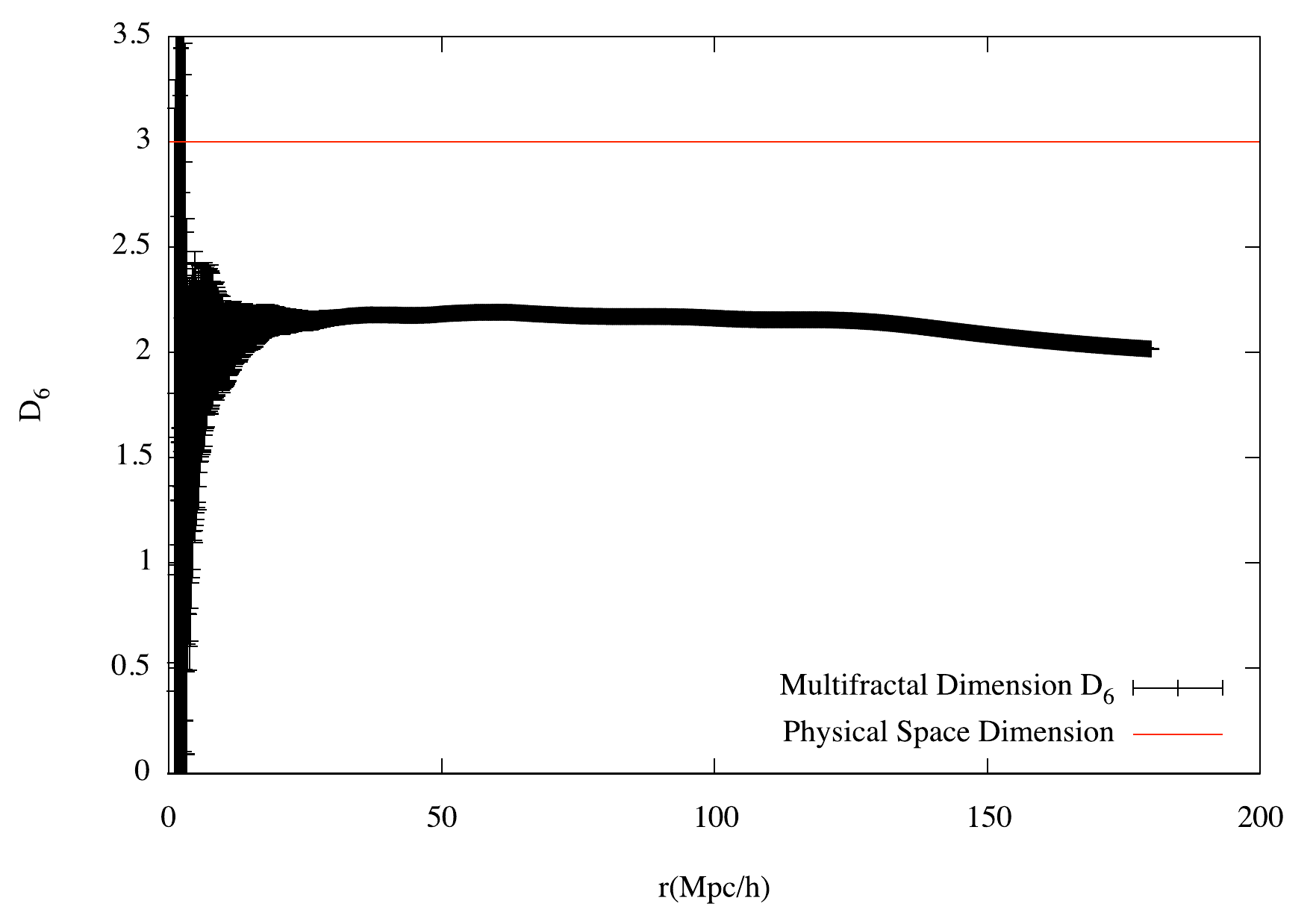}
	
	\caption{Multifractal spectrum  $D_q(r)$ as a function of the radial distance $r$ for low-density environments, with $-6 \leqslant q  \leqslant 0$, corresponding to the galaxy clustering in SDSS DR9. Error bars, corresponding to $\pm1\sigma$  are indicated in the plots.} 
	
\label{Dqr2a}
\end{figure*}

For high-density regions, namely for the correlation integral for structure parameter values ??in the range $q \geqslant 1$, the dependence of the spectrum of multifractal dimension on radial distance is shown in Figure  \ref{Dqr2a}. 

\begin{figure*}
	\includegraphics[width=0.45\linewidth,clip]{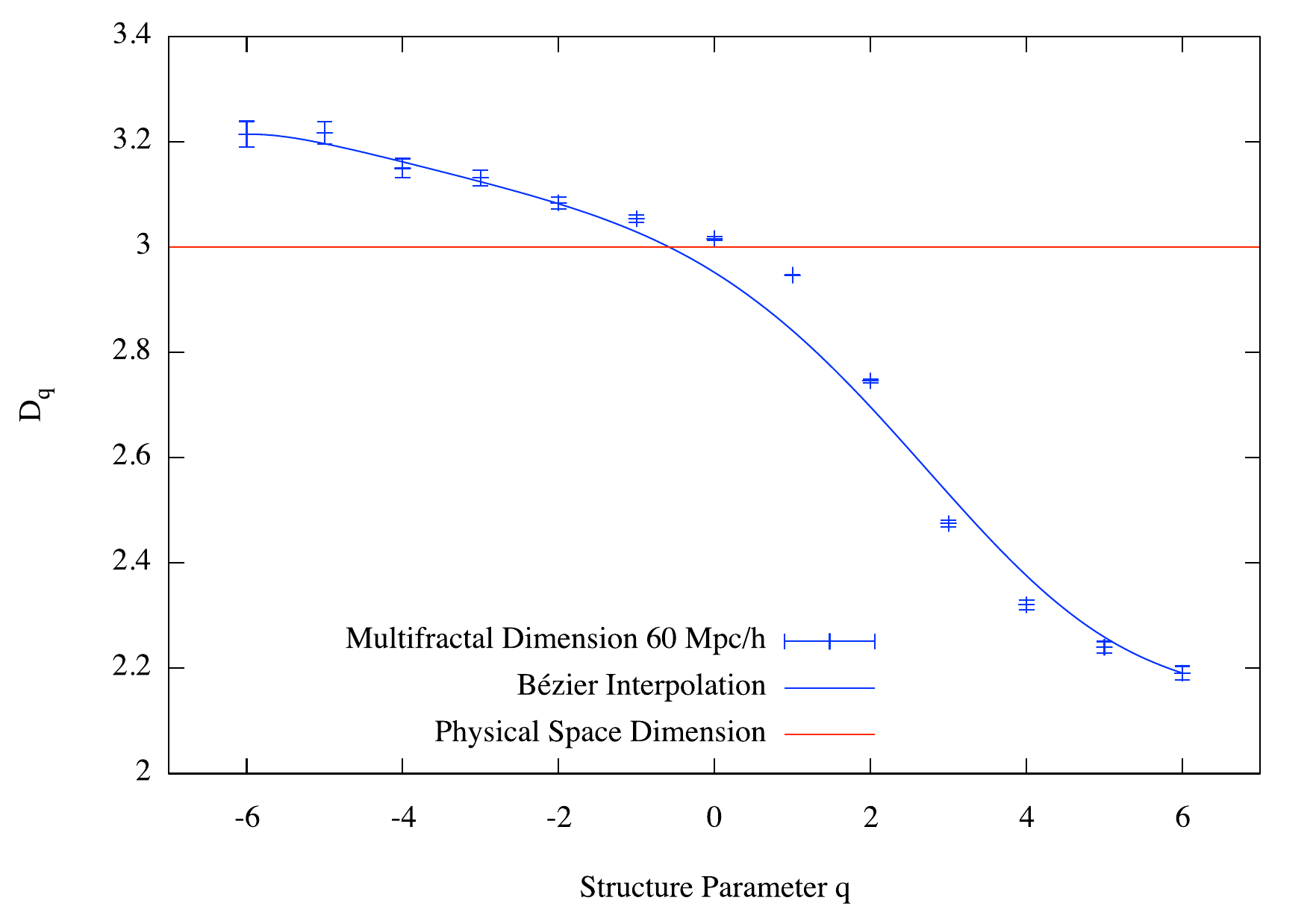}
	\includegraphics[width=0.45\linewidth,clip]{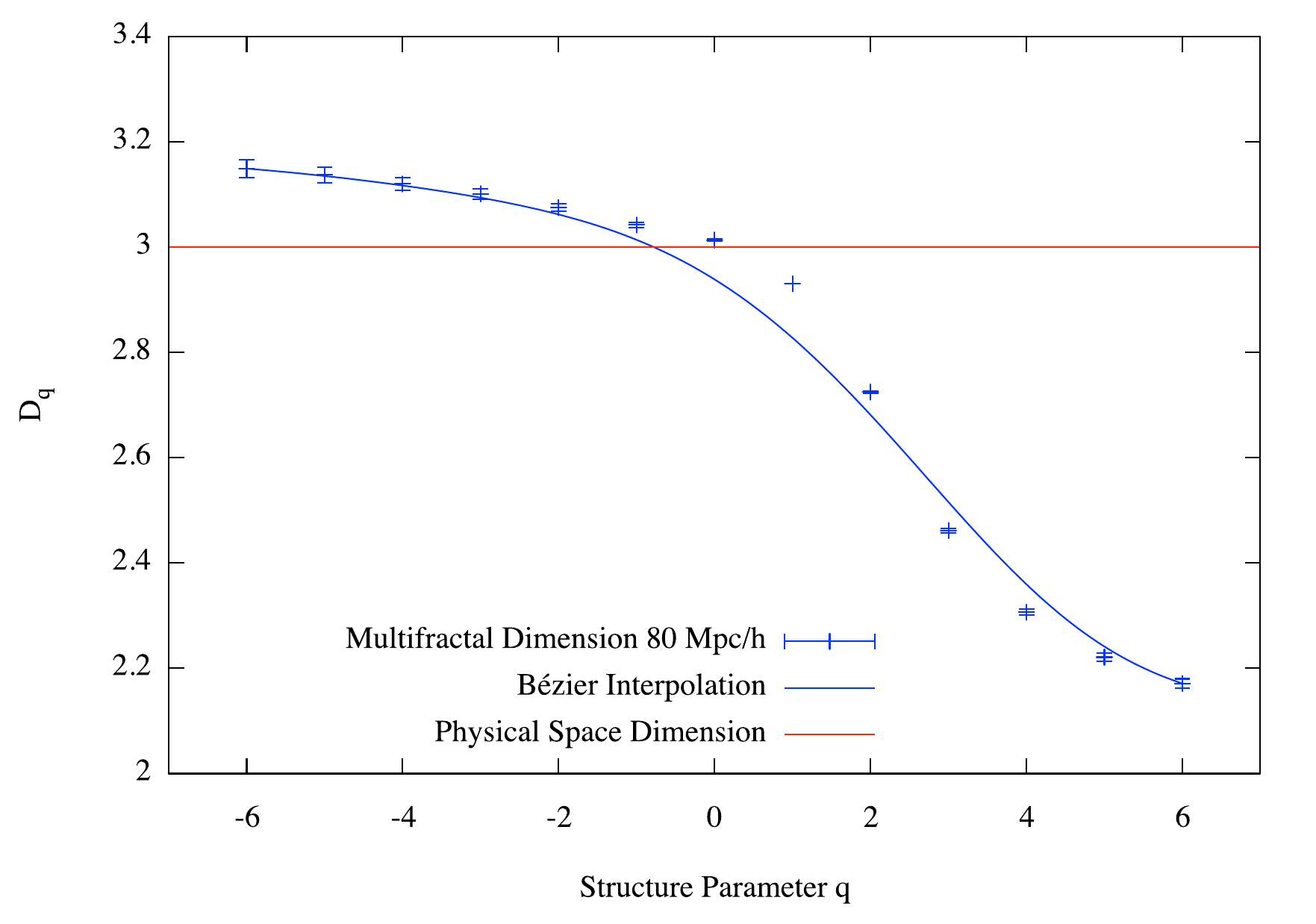}
	\includegraphics[width=0.45\linewidth,clip]{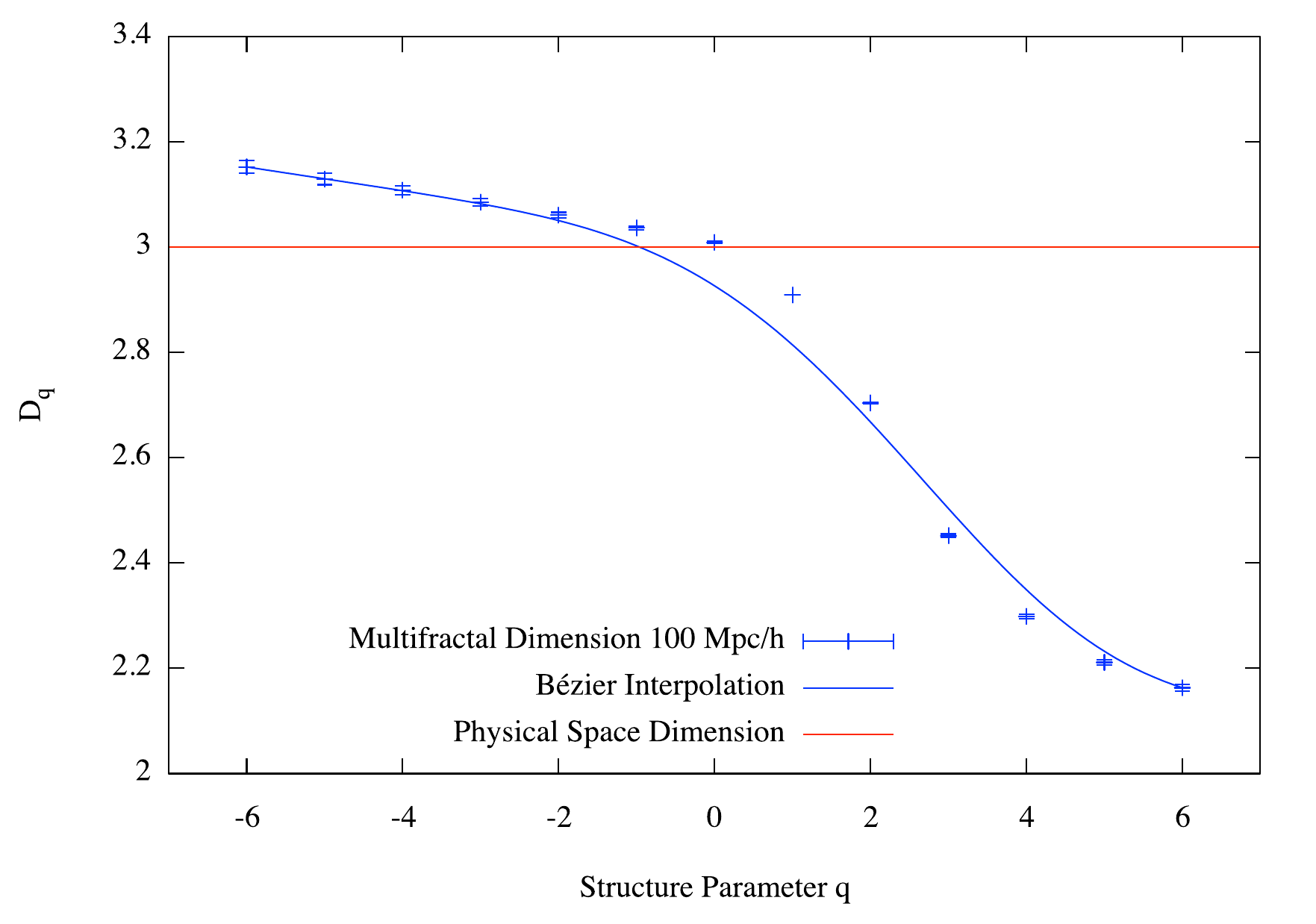}
         \includegraphics[width=0.45\linewidth,clip]{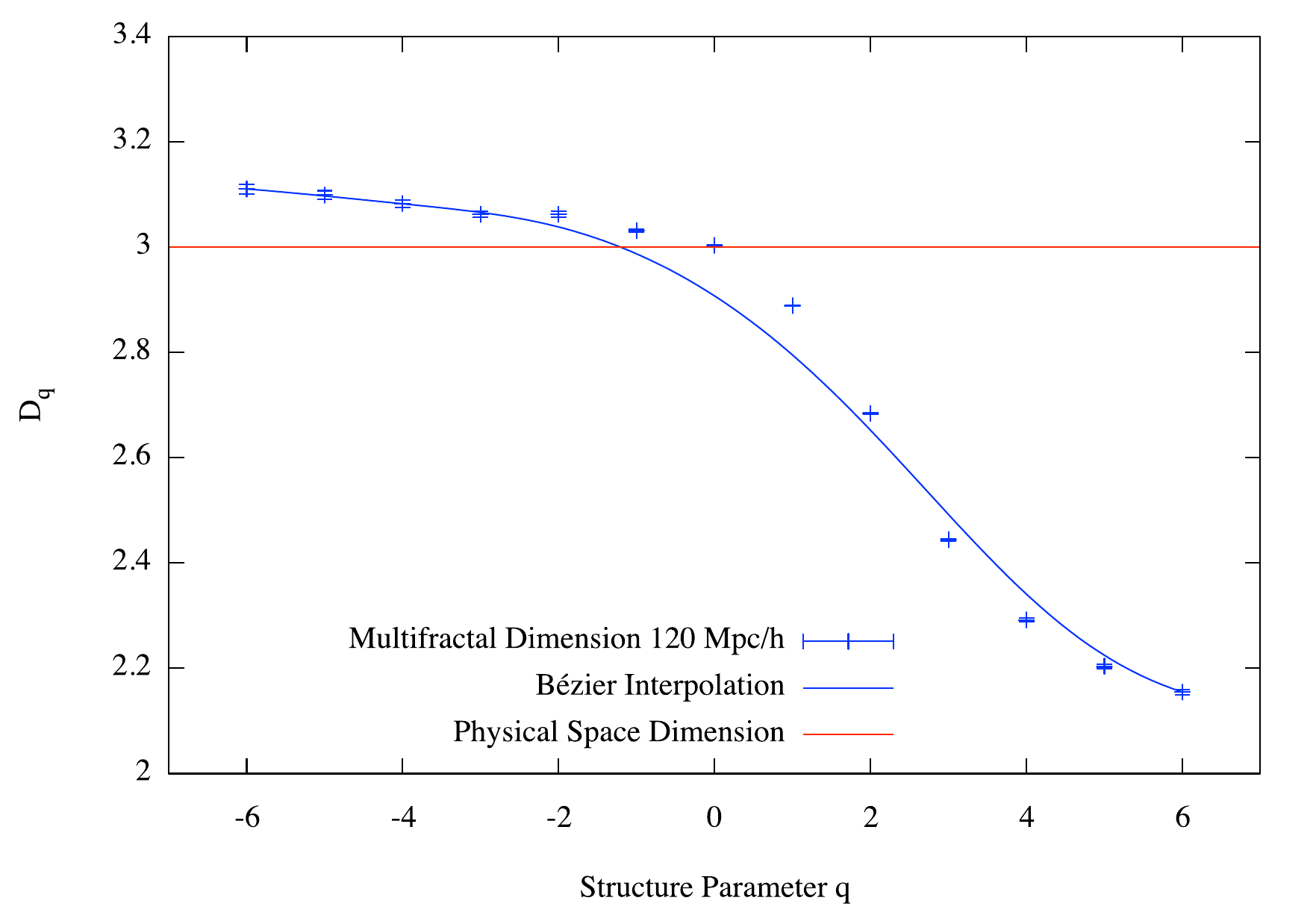}
         \includegraphics[width=0.45\linewidth,clip]{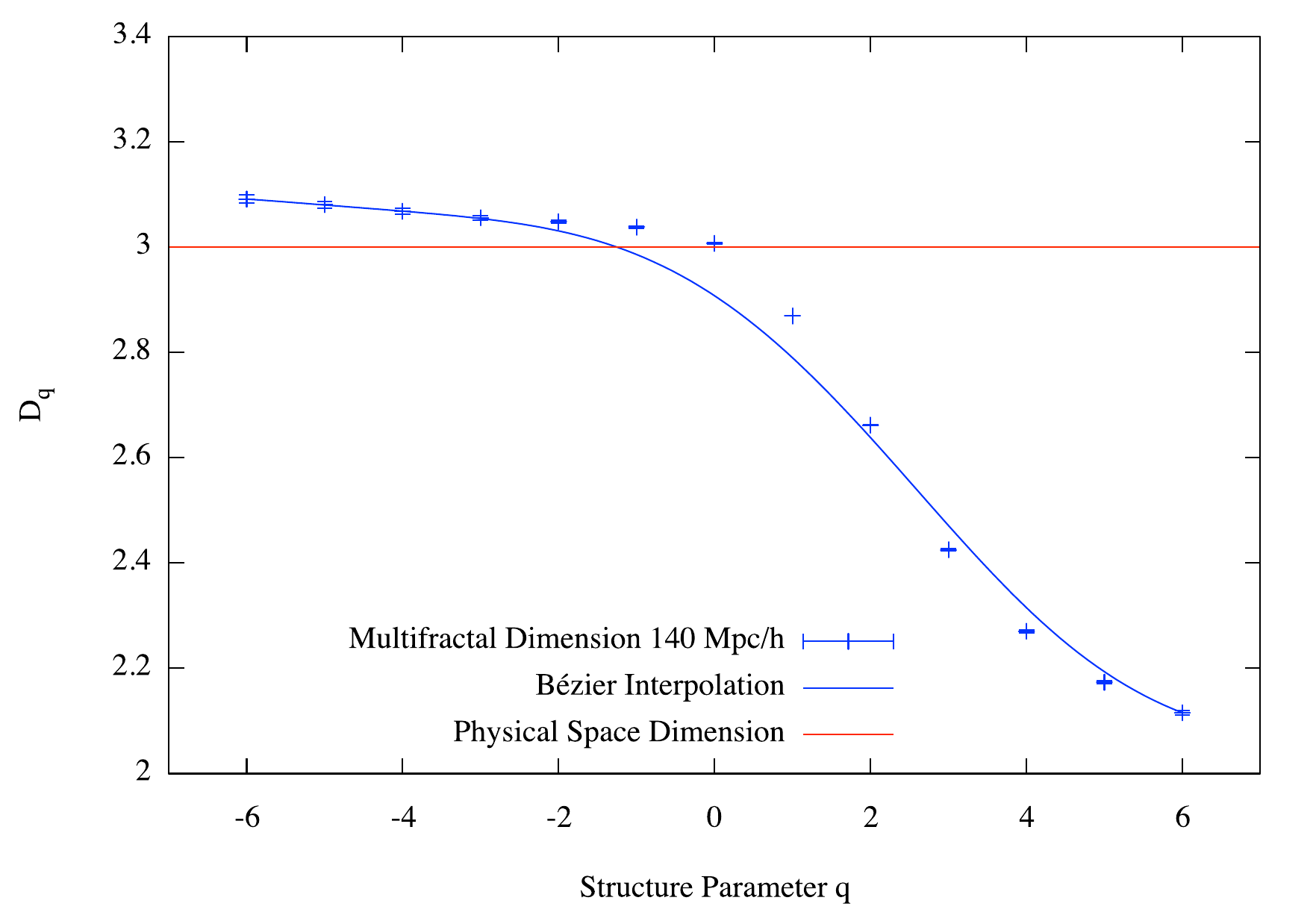}
         \includegraphics[width=0.45\linewidth,clip]{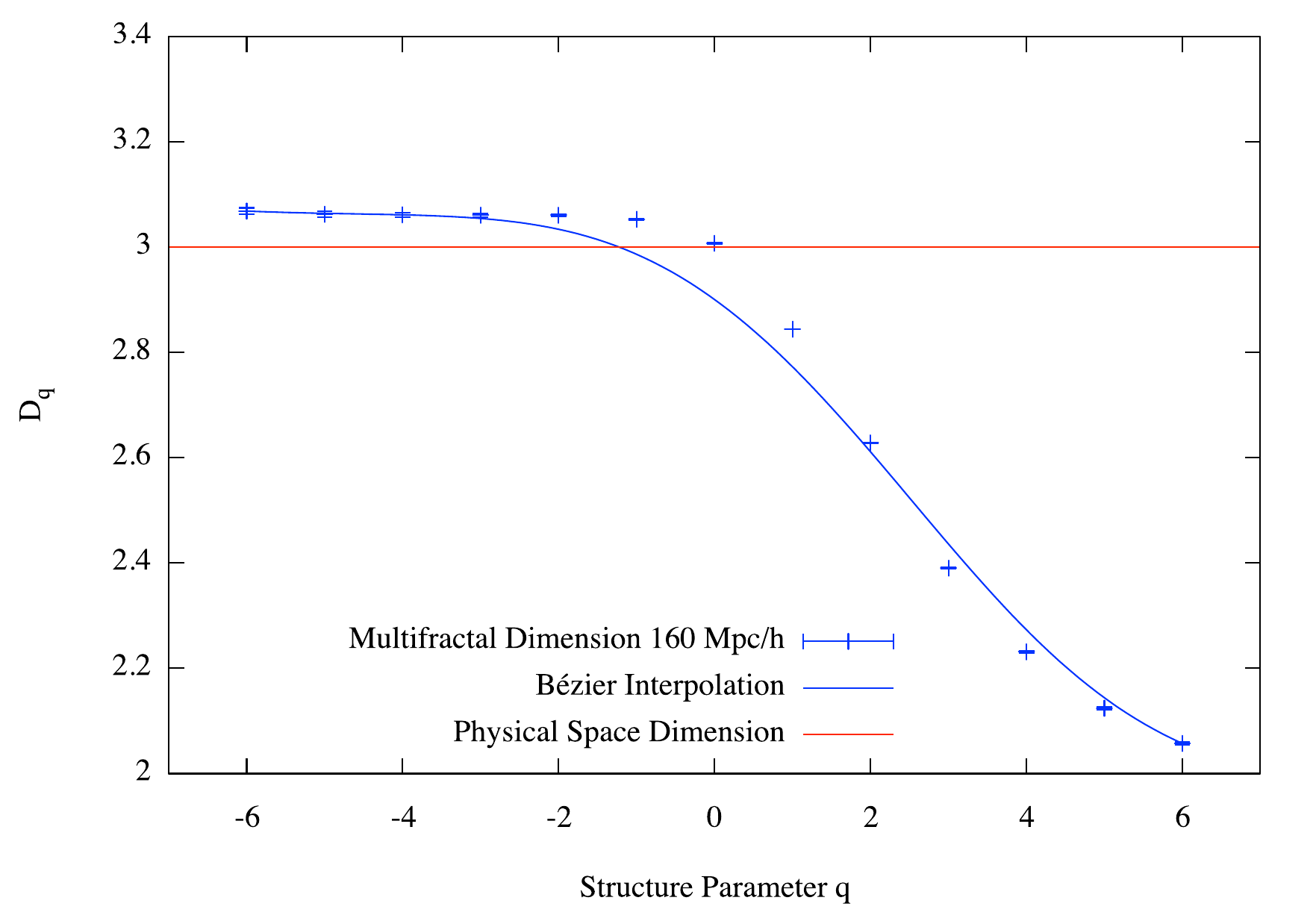}
         \includegraphics[width=0.45\linewidth,clip]{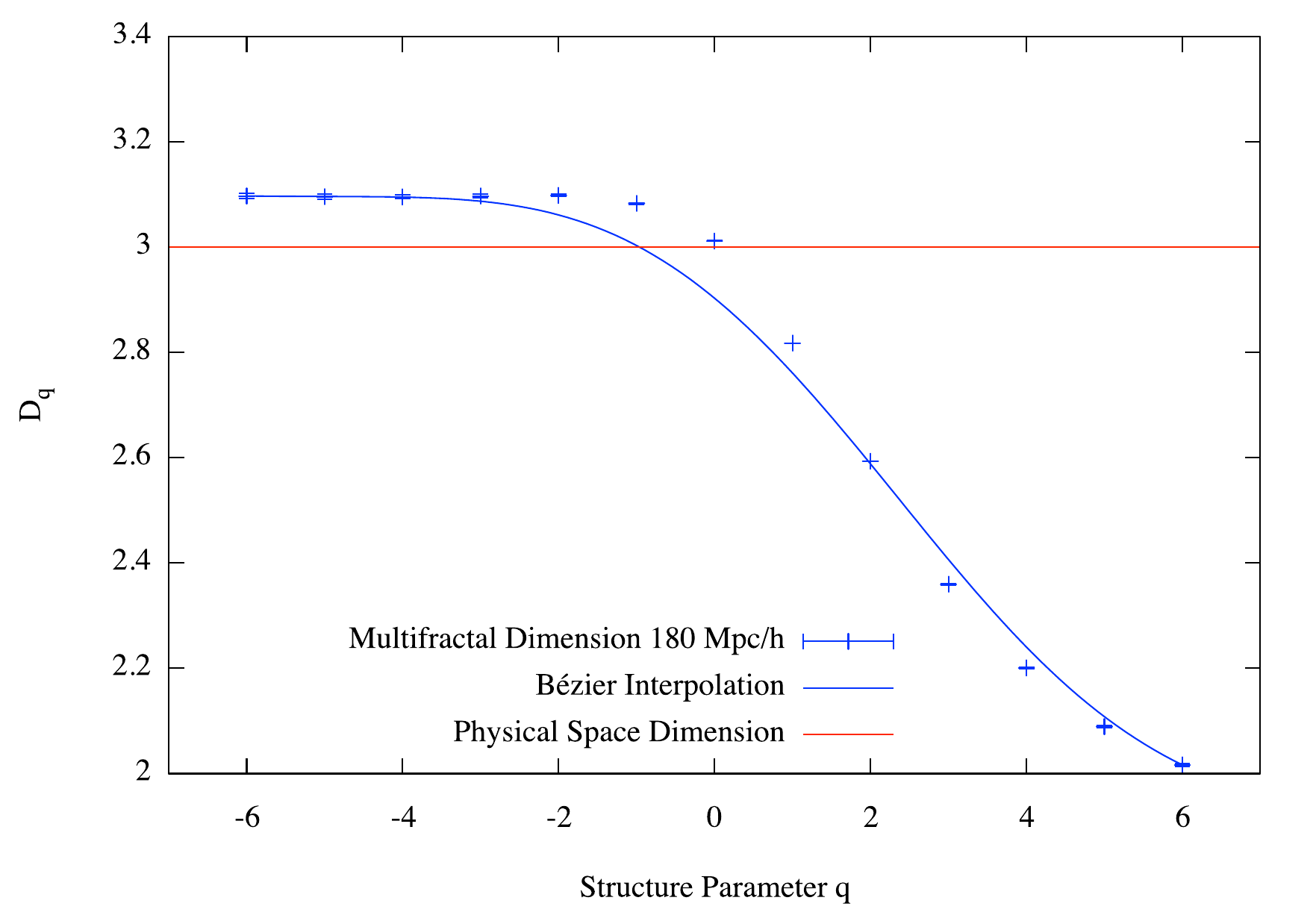} 
    
         \caption{Multi-fractal dimension spectrum $D_q(r)$ of the galaxy clustering  as a function of the structure parameter $q$ in the range $[6, 6]$. Multi-fractal behaviour is observed even at depths on the order of 180 Mpc/h around the centres. The solid line indicates the B\'ezier interpolation curve.}
	\label{Dqrq}
\end{figure*}

These results can be combined to determine the multifractal dimension spectrum as a function of the structure parameter $q$ for different radial distance scales, as shown in Figure \ref{Dqrq}

\begin{figure*}
	\includegraphics[width=0.45\linewidth,clip]{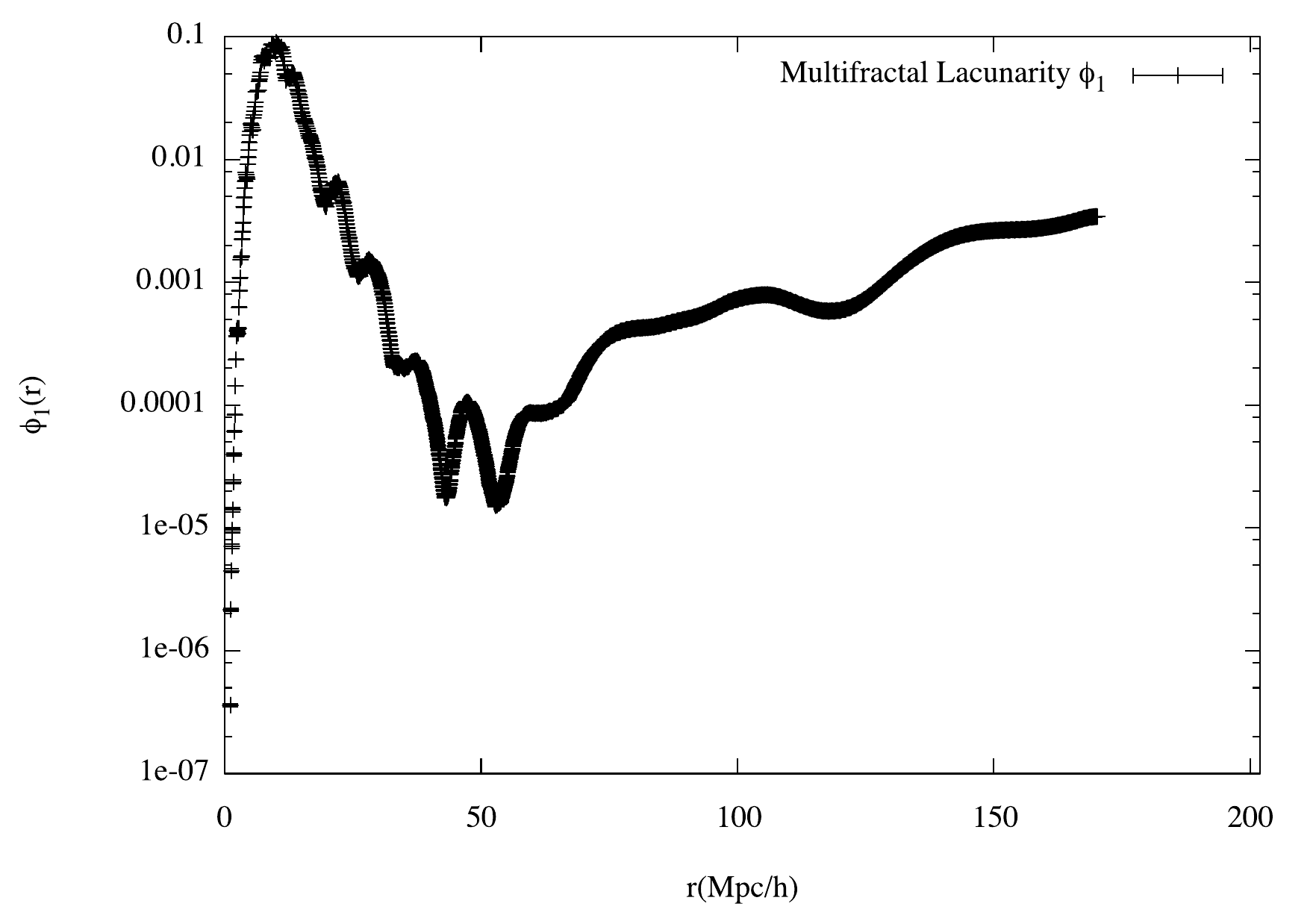}
	\includegraphics[width=0.45\linewidth,clip]{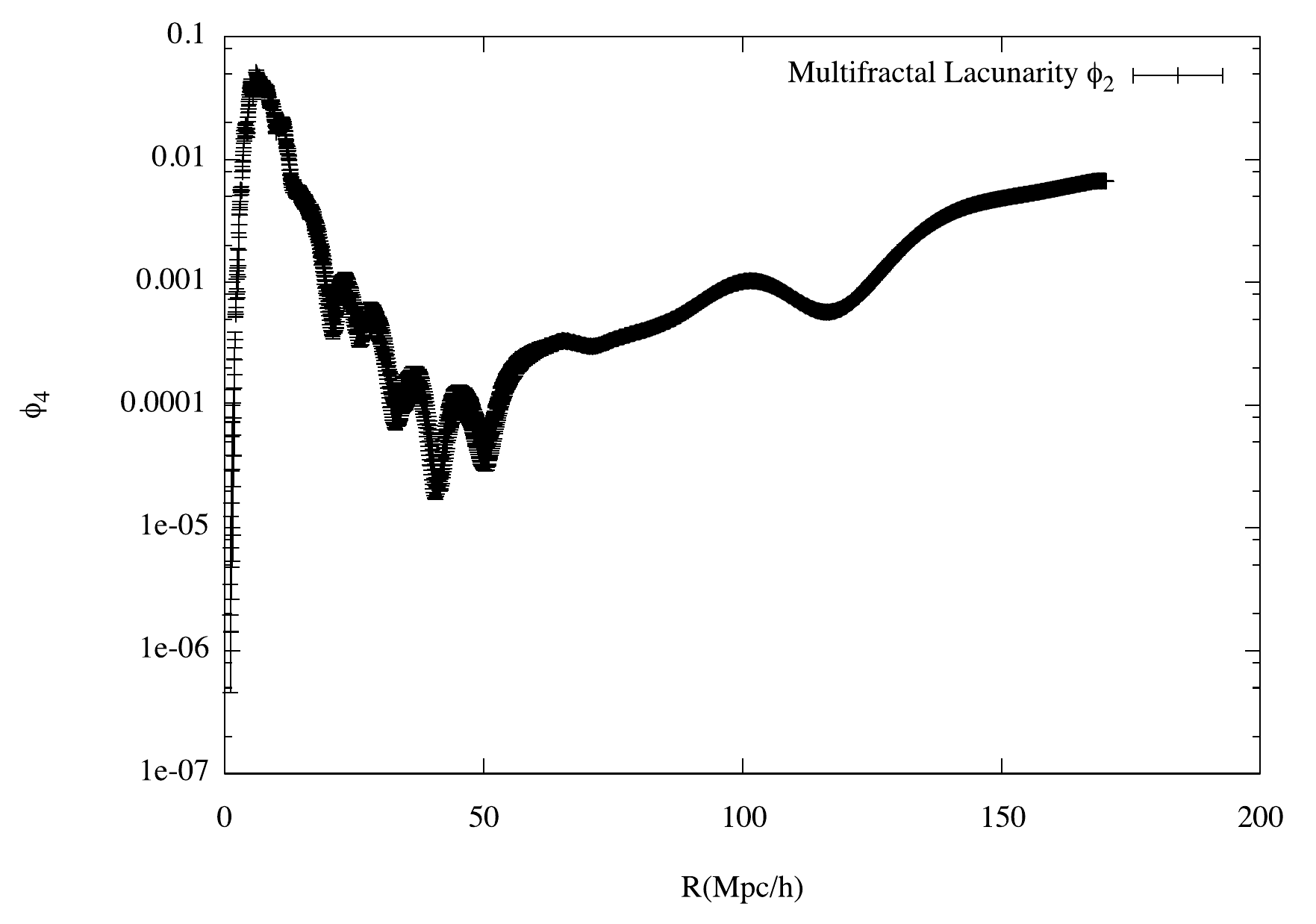}
	\includegraphics[width=0.45\linewidth,clip]{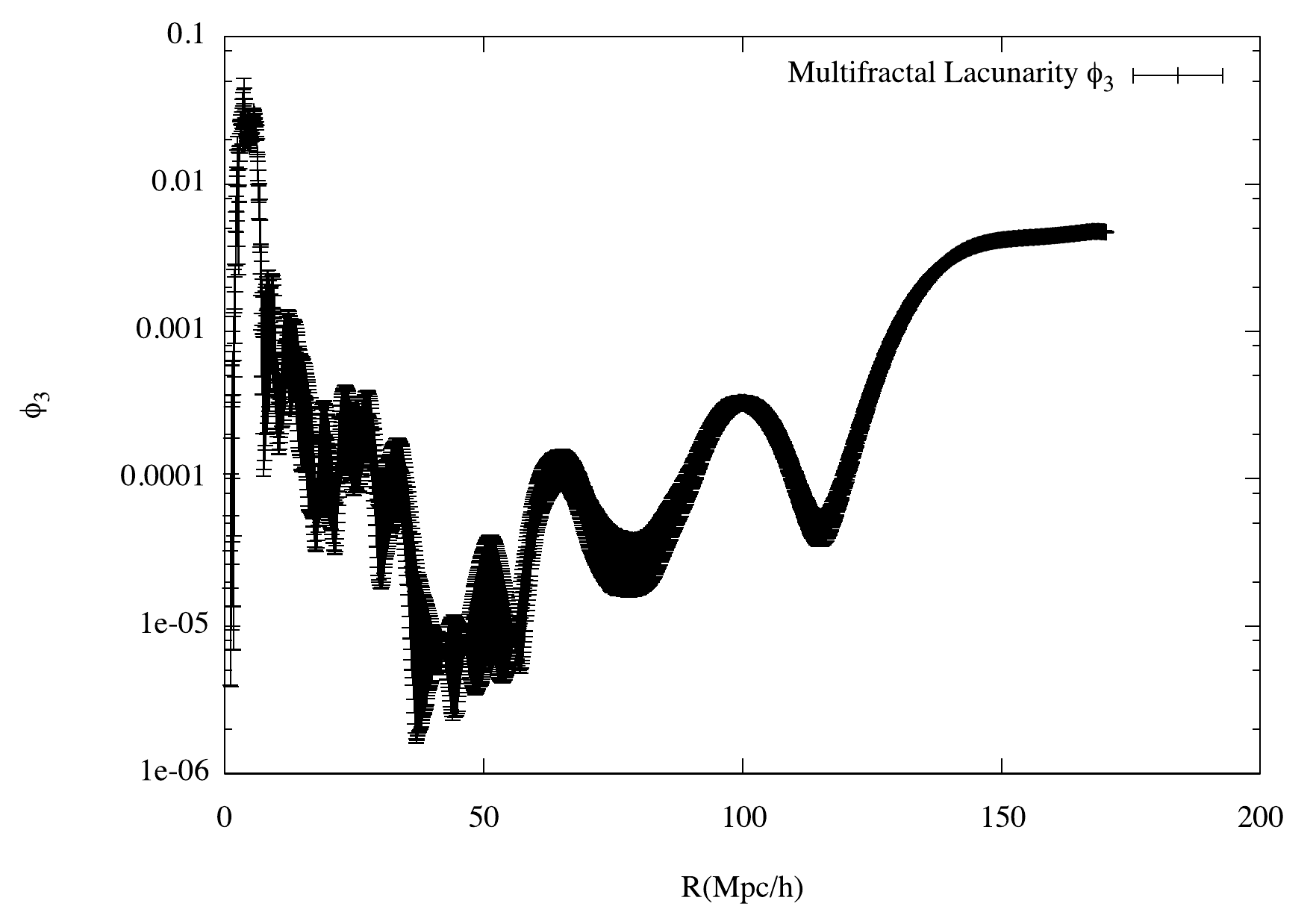}
	\includegraphics[width=0.45\linewidth,clip]{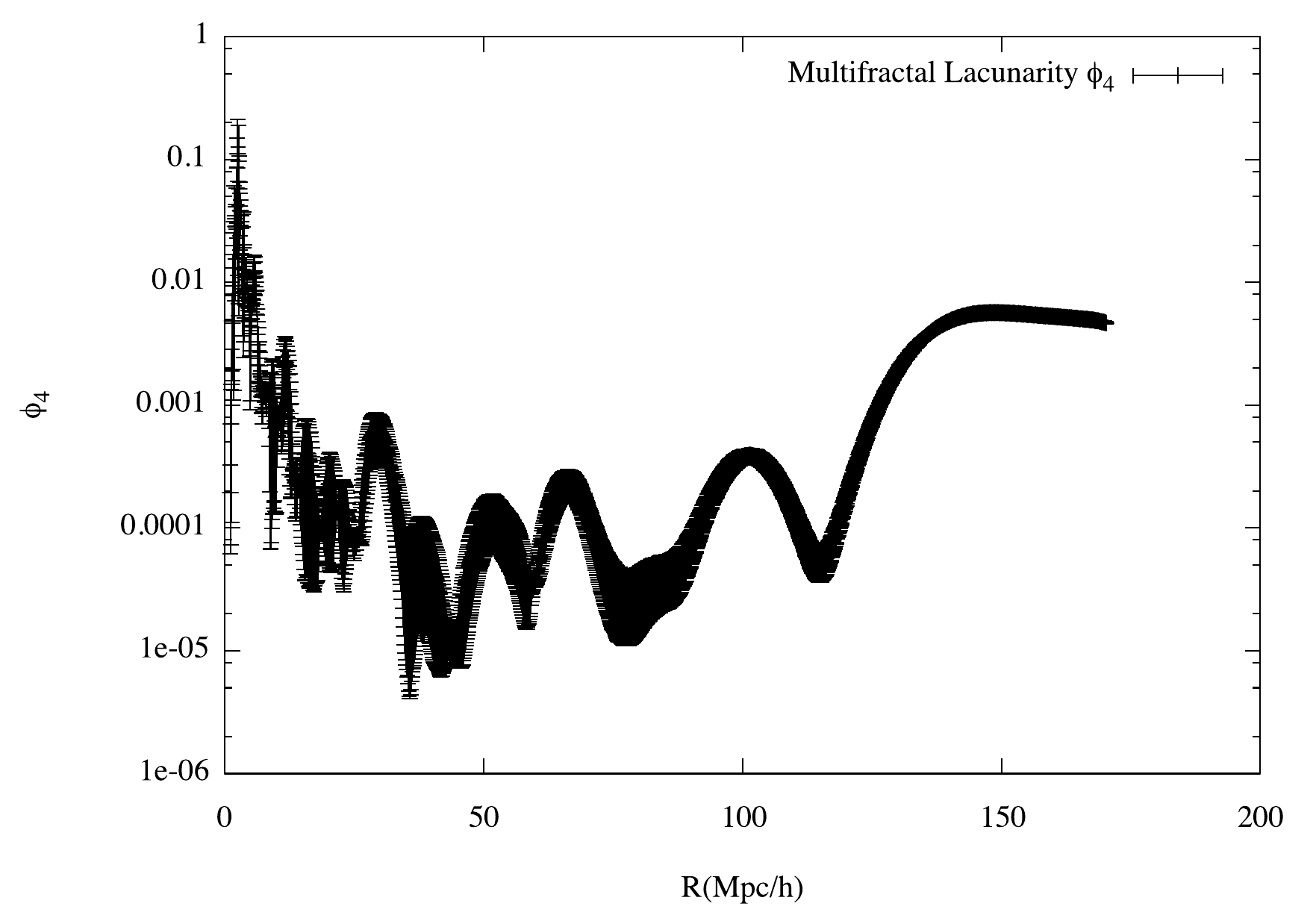}
         \includegraphics[width=0.45\linewidth,clip]{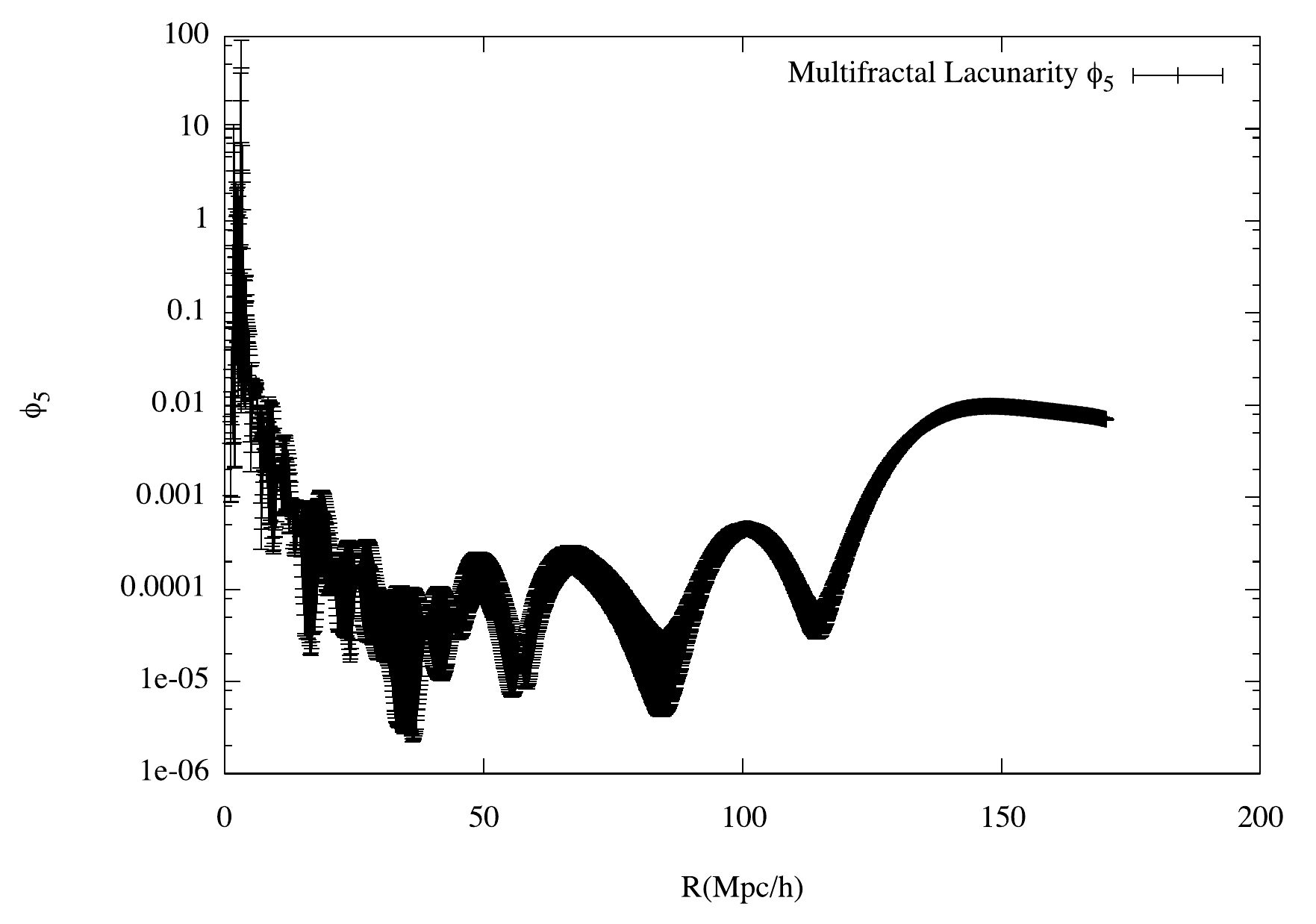}
         \includegraphics[width=0.45\linewidth,clip]{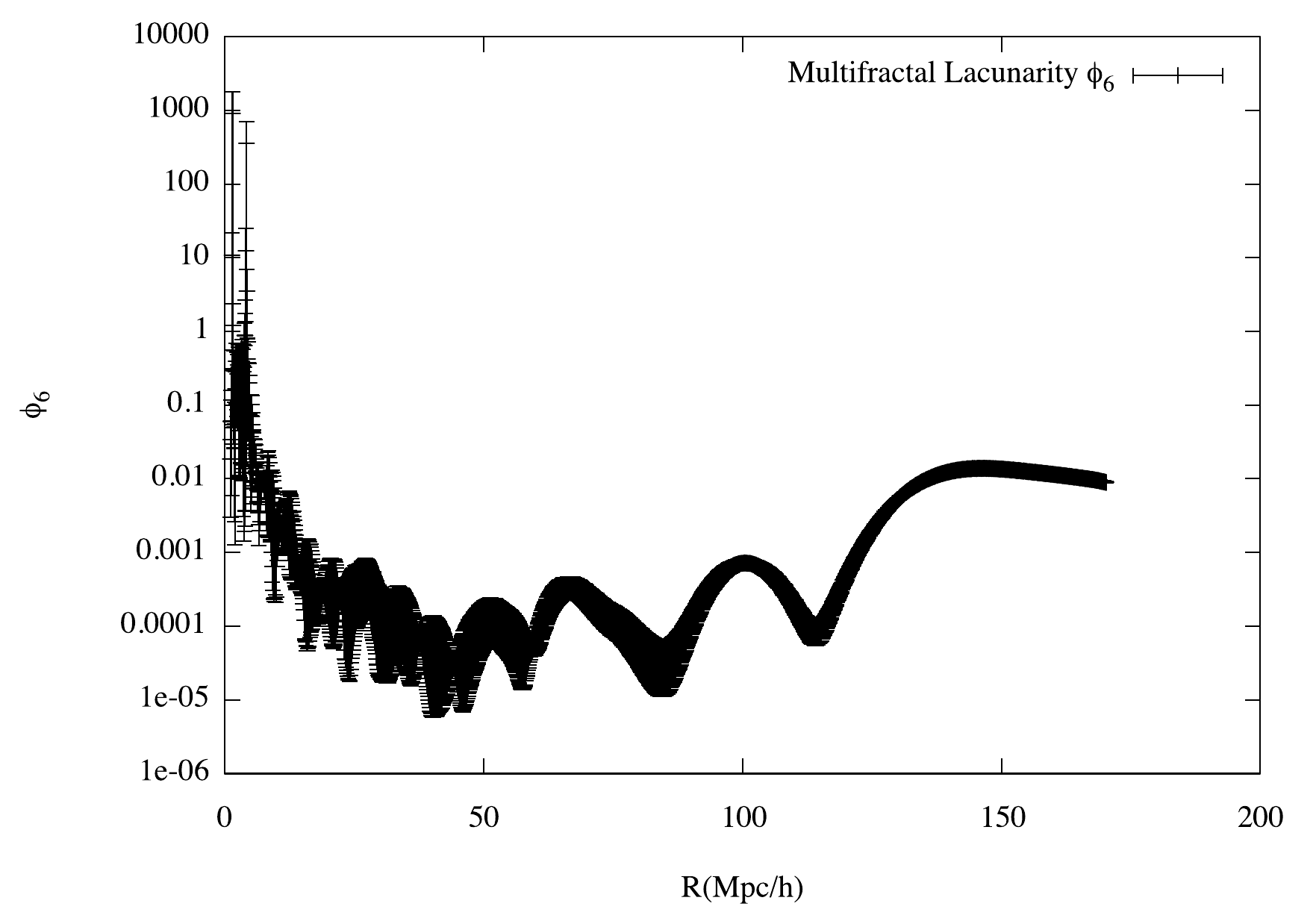} 
         \caption{Lacunarity spectrum $\Phi_q(r)$ in losgscale as function of the radial distance $r$ for high-density environments, with $q \geqslant 1$, according to SDSS (DR9). Error bars of $ \pm1\sigma$ are indicated in the graphs.}	
         \label{Phiqr}
\end{figure*}

Given that the fractal dimension does not specify how galaxies fill space, our analysis is completed by the calculation of the lacunarity spectrum for the same values of the structure parameter. Figure \ref{Phiqr}  shows the corresponding results. 

\section{Discussion}

 \citet{2012arXiv1209.2637C} have performed fractal analysis on the clustering of dark matter halos from the Millennium Simulation \citep{2005Natur.435..629S}. We successfully applied the same methodology to the clustering of galaxies in DR9 for the local Universe, i.e., galaxies for which the Hubble distance can be used for the localisation of points in the set without using an a priori cosmological model. The analysed sample was obtained from the most complete and connected region within the galactic catalogue (with right ascension in the range  $120 < \alpha < 240 $ and declination in the range $0 < \delta < 60 $ ). The sample is a magnitude-limited sample of 164,168 galaxies with redshifts in the range $2.3\times10^{-3} < z < 0.15$. The calculations performed accounted for the effective depth for which the fractal behaviour could be analysed, up to a maximum radial distance of 202 Mpc/h, which is the maximum radius of the sphere that we can inscribe within the limits of our sample of galaxies without introducing bias for subsequent statistical analysis \citep{citeulike:2502783}.

We directly calculated the multifractal dimension using the correlation integral. The generalised correlation integral found for galaxy clustering behaves as expected and is in agreement with the results of \citep{2012arXiv1209.2637C} and \citep{2009MNRAS.399L.128S}. The resulting behaviour corresponds to an increase in the average number of neighbours that are located around the centres in high-density regions, for which $q \geqslant 1$, and a decrease in the number of neighbours for low-density regions (voids with $q < 1$).

Using the sliding-window technique, the fractal dimension was calculated for low-density regions with structural parameter values $q < 1$. In agreement with the findings for the Millennium Simulation \citep{2012arXiv1209.2637C},  we found a growth that surpasses the physical dimension of space in regions near the centres, followed by a rapid tendency toward homogeneity at radial distances greater than 50 Mpc/h. In contrast, for high-density regions with structure parameter values $q \geqslant 1$, the behaviour differs from that we found previously \citep{2012arXiv1209.2637C}. Although there is a region of rapid dimensional growth, asymptotic behaviour toward the physical dimension of space is not observed. In our case, the dimension reaches a high value in regions around the centres, where the tendency towards homogeneity is observed, followed by a second region where the dimension of the galaxy clustering tends to differ from the physical dimension of space, thus showing persistent fractal behaviour at larger scales. This is consistent with galaxy clustering behaviour on large scales, as published by other research groups \citet{2005A&A...443...11J}, \citet{2011ARep...55..324V}, It is also consistent with analysis of the clustering of quasars \citet{2012arXiv1201.5554R} and with studies of the anisotropy of the cosmic background radiation  \citet{2011JPSJ...80g4003K}.  

The determination of the lacunarity spectrum is spoiled by the propagation of errors inherent in its calculation based on the pre-factor F. Notwithstanding this difficulty, oscillatory behaviour (voids alternating with regions of high concentration of galaxies) can be detected, as was found in the Millennium Simulation, with a lacunarity maximum in regions below 10 Mpc/h. However, in our case, monotonically decreasing behaviour was not observed. This result is consistent with deviations from the transition to homogeneity detected by the multifractal-dimension set because the lacunarity values increase again after 50 Mpc/h.

\section{Conclusions}

In this article, we analysed galaxy clustering in the local Universe from the multifractal point of view. Our results led us to conclude the following:

No transition to homogeneity was observed in galaxy clustering, at least for distance scales less than approximately 180 Mpc/h; instead, there was a clear persistence of multifractal behaviour. For values of the structure parameter corresponding to low densities, the calculated dimensions exhibited an unequivocal tendency to fill space almost homogeneously at scales greater than 50 Mpc/h. However, for galaxy clustering, the fractal-dimension spectrum in all cases showed two clearly defined regions: a first region of high-dimensional growth on scales less than 30 Mpc/h and a second region of smoothly decaying structure with fractal dimension below that of physical space at depths greater than approximately 30 Mpc/h. Therefore, the use of multifractal analysis is most suitable for the characterisation of galaxy clustering.

The lacunarity spectrum confirmed the persistence of the fractal behaviour of the set, showing oscillatory behaviour with alternation between high and low lacunarity values, i.e., the presence of voids followed by regions of high galaxy clustering. We did not observe a lacunarity decrease that would otherwise allow detecting homogeneity. This result confirms the departure from homogeneity within the analysed volume.

The sample is eminently multifractal. A tendency towards monofractality (i.e., only one fractal-dimension value independent of the structure parameter) was not observed in the galactic sample for the distance scale of this sample. This result implies that further detailed research is needed to investigate the standard cosmological principle and its consistency with observations. Although we cannot conclude that a transition to homogeneity cannot occur at greater depths, the problem of introducing homogenisation due to cosmological assumptions should be avoided, with the goal of approaching this problem without introducing biases in the FRW metric.

\section*{Acknowledgements}
C. A. Chac\'on-Cardona thanks the Universidad Distrital  Francisco Jos\'e de Caldas for their financial support to perform doctoral studies. 
  
\bibliography{my_citations}

\begin{thebibliography}{26}
\expandafter\ifx\csname natexlab\endcsname\relax\def\natexlab#1{#1}\fi

\bibitem[{{Bagla}, {Yadav} \& {Seshadri}(2008){Bagla}, {Yadav}, \&
  {Seshadri}}]{2008MNRAS.390..829B}
{Bagla} J.~S., {Yadav} J., {Seshadri} T.~R., 2008, \mnras, 390, 829

\bibitem[{Blumenfeld \& Mandelbrot(1997)}]{PhysRevE.56.112}
Blumenfeld R., Mandelbrot B.~B., 1997, Phys. Rev. E, 56, 112

\bibitem[{{Chac{\'o}n-Cardona} \& {Casas-Miranda}(2012)}]{2012arXiv1209.2637C}
{Chac{\'o}n-Cardona} C.~A., {Casas-Miranda} R.~A., 2012, \mnras, 427, 2613

\bibitem[{{Enqvist}(2008)}]{2008GReGr..40..451E}
{Enqvist} K., 2008, General Relativity and Gravitation, 40, 451

\bibitem[{Gabrielli {et~al}\mbox{.}(2005)Gabrielli, Sylos, Joyce, \&
  Pietronero}]{citeulike:2502783}
Gabrielli A., Sylos, Joyce M., Pietronero L., 2005, {Statistical Physics for
  Cosmic Structures}. Springer Verlag

\bibitem[{{Gaite}(2007)}]{2007ApJ...658...11G}
{Gaite} J., 2007, \apj, 658, 11

\bibitem[{{Grujic} \& {Pankovic}(2009)}]{2009arXiv0907.2127G}
{Grujic} P., {Pankovic} V., 2009, ArXiv e-prints

\bibitem[{Hausdorff(1918)}]{springerlink:10.1007/BF01457179}
Hausdorff F., 1918, Mathematische Annalen, 79, 157, 10.1007/BF01457179

\bibitem[{{Joyce} {et~al}\mbox{.}(2005){Joyce}, {Sylos Labini}, {Gabrielli},
  {Montuori}, \& {Pietronero}}]{2005A&A...443...11J}
{Joyce} M., {Sylos Labini} F., {Gabrielli} A., {Montuori} M., {Pietronero} L.,
  2005, \aap, 443, 11

\bibitem[{{Kobayashi} {et~al}\mbox{.}(2011){Kobayashi}, {Yamazaki}, {Kuninaka},
  {Katori}, {Matsushita}, {Matsushita}, \& {Chiang}}]{2011JPSJ...80g4003K}
{Kobayashi} N., {Yamazaki} Y., {Kuninaka} H., {Katori} M., {Matsushita} M.,
  {Matsushita} S., {Chiang} L.-Y., 2011, Journal of the Physical Society of
  Japan, 80, 074003

\bibitem[{Longair(2008)}]{Longair:2008}
Longair M., 2008, Galaxy formation, Astronomy and astrophysics library.
  Springer

\bibitem[{Mandelbrot(1983)}]{mandelbrot1983fractal}
Mandelbrot B., 1983, The fractal geometry of nature. W.H. Freeman

\bibitem[{Mart{\'\i}nez \& Saar(2002)}]{martinez2002statistics}
Mart{\'\i}nez V., Saar E., 2002, Statistics of the galaxy distribution. Chapman
  \& Hall/CRC

\bibitem[{{Nakamichi} \& {Morikawa}(2004)}]{2004PhyA..341..215N}
{Nakamichi} A., {Morikawa} M., 2004, Physica A Statistical Mechanics and its
  Applications, 341, 215

\bibitem[{Peacock(1999)}]{peacock1999cosmological}
Peacock J., 1999, Cosmological physics. Cambridge University Press

\bibitem[{{Provenzale}, {Spiegel} \& {Thieberger}(1997){Provenzale}, {Spiegel},
  \& {Thieberger}}]{1997Chaos...7...82P}
{Provenzale} A., {Spiegel} E.~A., {Thieberger} R., 1997, Chaos, 7, 82

\bibitem[{{Ribeiro}(1992)}]{1992ApJ...388....1R}
{Ribeiro} M.~B., 1992, \apj, 388, 1

\bibitem[{{Ribeiro}(2005)}]{2005A&A...429...65R}
---, 2005, \aap, 429, 65

\bibitem[{{Rozgacheva} {et~al}\mbox{.}(2012){Rozgacheva}, {Borisov}, {Agapov},
  {Pozdneev}, \& {Shchetinina}}]{2012arXiv1201.5554R}
{Rozgacheva} I.~K., {Borisov} A.~A., {Agapov} A.~A., {Pozdneev} I.~A.,
  {Shchetinina} O.~A., 2012, ArXiv e-prints

\bibitem[{{Sarkar} {et~al}\mbox{.}(2009){Sarkar}, {Yadav}, {Pandey}, \&
  {Bharadwaj}}]{2009MNRAS.399L.128S}
{Sarkar} P., {Yadav} J., {Pandey} B., {Bharadwaj} S., 2009, \mnras, 399, L128

\bibitem[{{Schlegel}, {Finkbeiner} \& {Davis}(1998){Schlegel}, {Finkbeiner}, \&
  {Davis}}]{1998ApJÉ500..525S}
{Schlegel} D.~J., {Finkbeiner} D.~P., {Davis} M., 1998, \apj, 500, 525

\bibitem[{Scrimgeour {et~al}\mbox{.}(2012)Scrimgeour, Davis, Blake, James,
  Poole, {et~al.}}]{Scrimgeour:2012wt}
Scrimgeour M., Davis T., Blake C., James J.~B., Poole G., {et~al.}, 2012,
  Mon.Not.Roy.Astron.Soc., 425, 116

\bibitem[{{SDSS-III Collaboration} {et~al}\mbox{.}(2012){SDSS-III
  Collaboration}, {:}, {Ahn}, {Alexandroff}, {Allende Prieto}, {Anderson},
  {Anderton}, {Andrews}, {Bailey}, {Barnes}, \& et~al.}]{2012arXiv1207.7137S}
{SDSS-III Collaboration} {et~al.}, 2012, ArXiv e-prints

\bibitem[{{Springel} {et~al}\mbox{.}(2005){Springel}, {White}, {Jenkins},
  {Frenk}, {Yoshida}, {Gao}, {Navarro}, {Thacker}, {Croton}, {Helly},
  {Peacock}, {Cole}, {Thomas}, {Couchman}, {Evrard}, {Colberg}, \&
  {Pearce}}]{2005Natur.435..629S}
{Springel} V. {et~al.}, 2005, \nat, 435, 629

\bibitem[{{Strauss} {et~al}\mbox{.}(2002){Strauss}, {Weinberg}, {Lupton},
  {Narayanan}, {Annis}, {Bernardi}, {Blanton}, {Burles}, {Connolly},
  {Dalcanton}, {Doi}, {Eisenstein}, {Frieman}, {Fukugita}, {Gunn},
  {Ivezi{\'c}}, {Kent}, {Kim}, {Knapp}, {Kron}, {Munn}, {Newberg}, {Nichol},
  {Okamura}, {Quinn}, {Richmond}, {Schlegel}, {Shimasaku}, {SubbaRao},
  {Szalay}, {Vanden Berk}, {Vogeley}, {Yanny}, {Yasuda}, {York}, \&
  {Zehavi}}]{2002AJ....124.1810S}
{Strauss} M.~A. {et~al.}, 2002, \aj, 124, 1810

\bibitem[{{Verevkin}, {Bukhmastova} \& {Baryshev}(2011){Verevkin},
  {Bukhmastova}, \& {Baryshev}}]{2011ARep...55..324V}
{Verevkin} A.~O., {Bukhmastova} Y.~L., {Baryshev} Y.~V., 2011, Astronomy
  Reports, 55, 324

\end{thebibliography}

\label{lastpage}

\end{document}